\documentclass{aa}  

\usepackage{graphicx}
\usepackage{supertabular,booktabs}
\usepackage{txfonts}
\usepackage{graphicx}
\usepackage{longtable}
\usepackage{xcolor}
\usepackage{txfonts}
\usepackage{natbib,twoopt}
\usepackage[breaklinks=true]{hyperref} %% to avoid \citepads line fills
\bibpunct{(}{)}{;}{a}{}{,}             %% natbib format for A&A and ApJ
\usepackage{placeins}
\usepackage{subfigure}

\begin{document}

   \title{Survey of Orion Disks with ALMA (SODA)}

   \subtitle{III: Disks in wide binary systems in L1641 and L1647}

   \author{G. Ricciardi\inst{1,} \inst{2,} \inst{3}
          \and
          S.E. van Terwisga\inst{2,} \inst{4}
          \and
          V. Roccatagliata\inst{5,} \inst{6}
          \and
          A. Hacar\inst{7}
          \and
          T. Henning\inst{2}
          \and
          W. Del Pozzo\inst{3,} \inst{8}
          }

   \institute{European Southern Observatory, Karl-Schwarzschild-Str2, 85748                Garching, Germany 
         \and
         Max-Planck-Institut für Astronomie, Königstuhl 17, 69117 Heidelberg, Germany 
         \and
         Università di Pisa, Physics Department E. Fermi, Largo Bruno Pontecorvo 3, 56127 Pisa, Italy
         \and
         Space Research Institute, Austrian Academy of Sciences, Schmiedlstr. 6, A-8042, Graz, Austria
         \and
         Alma Mater Studiorum, Università di Bologna, Dipartimento di Fisica e Astronomia (DIFA), Via Gobetti 93/2, 40129 Bologna, Italy
         \and
         INAF-Osservatorio Astrofisico di Arcetri, Largo E. Fermi 5, 50125 Firenze, Italy
         \and
        Department of Astrophysics, University of Vienna, Türkenschanzstraße 17 (Sternwarte) 1180 Wien, Austria
         \and
         INFN, Sezione di Pisa, Largo Bruno Pontecorvo 3, 56127 Pisa, Italy
             }

   \date{\today}

  \abstract
   {Observations of protoplanetary disks within multiple systems in nearby Star-Forming Regions (SFRs) have shown that the presence of a neighboring object influences the evolution of dust in disks. However, the sample size available and the separation range covered are insufficient to fully understand the dust evolution in binary systems.}
   {The goal of this work is to comprehensively characterize the impact of stellar multiplicity on Class~II disks in the L1641 and L1647 regions of Orion~A ($\sim$ 1-3~Myr), part of the Survey of Orion Disks with ALMA (SODA). We characterize the protostellar multiplicity using the Atacama Large Millimeter/submillimeter Array (ALMA), the ESO-VISTA, and Hubble Space telescopes. The resulting sample of 65 multiple systems represents the largest catalog of wide binary systems to date (projected separation $\geq$~1000~AU), allowing a more robust statistical characterization of the evolution and properties of protoplanetary disks.}
   {The disk population was observed in the continuum with ALMA at 225~GHz, with a median rms of 1.5~M$_{\oplus}$. Combining these data (resolution $\sim$1.1$^{''}$) with the ESO-VISTA near-infrared survey of the Orion~A cloud (resolution $\sim$0.7$^{''}$), multiple systems are assembled and selected by an iterative inside-out search in projected separation ($\geq$1000~AU).}
   {We identify 61 binary systems, 3 triple systems, and one quadruple system. The separation range is between 1000 and 10$^4$~AU. The dust mass distributions inferred with the Kaplan-Meier estimator yield a median mass of 3.23$^{+0.6}_{-0.4}$~M$_{\oplus}$ for primary disks and 3.88$^{+0.3}_{-0.3}$~M$_{\oplus}$ for secondary disks.} 
   {Combining our data with those available for Lupus and Taurus disks, we identify a threshold separation of about 130~AU, beyond which the previously observed positive correlation between millimeter flux (and hence dust mass) and projected separation is lost. Recent theoretical models confirm that pre- and post-threshold systems are the result of different star formation processes, such as the fragmentation of gravitationally unstable circumstellar disks, the thermal fragmentation of infalling cores, or the turbulent fragmentation of molecular clouds. We can rule out the dependency on different Star-Forming Regions: the cumulative mass distribution of multiples in SFRs of similar age are statistically indistinguishable. This result strengthens the hypothesis that there is a universal initial mass distribution for disks.}

   \keywords{protoplanetary disks - binaries: general – stars: pre-main sequence - methods: statistical}

   \maketitle
%
%-------------------------------------------------------------------

\section{Introduction}
Multiple systems are a common outcome of the star formation process \citep{Bate18, Duchene13, Larson01}, detected especially in the early stages \citep{Tobin16, Reipurth14}. The occurrence of binary systems is $\sim$50$\%$ in the local Solar neighborhood \citep{Raghavan10_SF}, the occurrence of multiple systems is about 30 - 50$\%$ in the field \citep{Raghavan10_SF} and up to about 70$\%$ in young clusters \citep{Kraus11}. Some studies even discuss the possibility that all stars may have formed in some kind of multiple configurations at the very beginning \citep{Kroupa95, Marks14}. In addition, multiplicity studies of pre-main sequence stars (Class II and Class III sources) have derived multiplicity fractions comparable to or exceeding those of main sequence stars \citep{Moe17, Kraus11, Reipurth07}. This prevalence across different class types and their close separation suggests that the origin of multiplicity is a direct consequence of the physical conditions of star formation. \par
Multiple star systems can be formed through various processes (and combinations thereof): (1) turbulent fragmentation of the molecular cloud \citep{Kawasaki23}, (2) thermal fragmentation of strongly perturbed, rotating, and infalling core \citep[e.g.,][]{Boss14, Boss13}, and/or (3) fragmentation of a gravitationally unstable circumstellar disk \citep[e.g.,][]{Longarini23, Stamatellos09, Machida08}. The first two scenarios will lead to multiples that are initially separated by several hundred to several thousand~AU, while scenario (3) leads to the formation of companions with separation of $\leq$100~AU \citep[e.g.,][]{Tobin13, Takakuwa12, Rodriguez98}. The separation of companions is therefore an important signature of the physics of star formation. \par
It is not clear how much the multiplicity of stars hinders planet formation and evolution. Numerical simulations have shown that in a given binary system there are some regions of orbit where planets are unstable \citep{Holman99}. Specifically, if the planet is orbiting one of the two stars, there is a maximum semi-major axis beyond which the planet's orbit is unstable; instead, if the planet is orbiting both stars, there is a minimum semi-major axis below which the planet is unstable.  Observations have shown that planets and protoplanetary disks easily form and survive around binary systems \citep{Duchene10, Bonavita07}. It should also be noted that in a star-forming environment, binaries have the added complication that they present a larger cross-section for encounters with other stars, and consequently the likelihood of interactions that could potentially destabilize the planetary system is higher than for systems orbiting a single star \citep{Adams06}. Therefore, it is crucial to better understand the effect of stellar multiplicity on the circumstellar disk structure, as this has a direct impact on the process of planet formation. \par
Ongoing improvements in observational techniques and hydrodynamical simulations have shed new light on our understanding of the evolution of isolated protoplanetary disks, such as the mechanisms responsible for mass accretion \citep{Espaillat22}, or the existence of multiphase instabilities and dynamical phenomena that can enhance planet formation rates \citep{Lesur22}. The presence of companion stars can significantly influence the dynamics and evolution of protoplanetary disks \citep{Pinte23}. On the one hand, companion stars can exert gravitational forces on the disk, leading to perturbations in the dust and gas distribution and in the orbital dynamics. On the other hand, tidal forces from the companions can induce gas flows or create regions of increased gas density, affecting the overall mass distribution and accretion rates within the disk, or affecting dust evolution processes such as grain growth and fragmentation. All of these aspects have implications for the subsequent evolution of the disk and the formation of planetary systems. \par
In the last decade, several observations of the gas and dust of disks in multiple star systems have been carried out, detecting compact ($<$ 100~AU) disks \citep[e.g.,][]{Rota22, Zagaria21, Zurlo21, Akeson19, Manara19, Cox17, Harris_2012}. The results confirm that the size and properties of the disks depend on the presence of companions. More recently, studies by \cite{Zagaria23}, \cite{Zhang23}, and \cite{Zagaria21} have shown that the evolution and lifetimes of disks around binary- or multiple- star systems differ from those of single stars. Nevertheless, the effects of stellar multiplicity on disk evolution are still poorly constrained: the available samples and the separation range do not allow for a complete picture of disk evolution. \par
In this paper, we focus on the dust to better constrain its evolution in disks in multiple star systems. Specifically, we use the Survey of Orion Disks with ALMA \footnote{https://emerge.univie.ac.at/results/soda-survey/} (SODA) (\citeauthor{vanTerwisga22}~2022), which with an angular resolution of $\sim$1.1$^{''}$ explores a broader disk population of 873 Class II disks around low-mass stars in Orion A below -6$^{\circ}$ declination. This survey is based on the Spitzer survey by \cite{Megeath12}, later updated to its final version by \cite{Megeath16}. \par
In order to well characterize the evolution of circumstellar dust around stars in multiple systems, we also include in our analysis the optical and near-infrared (NIR) data collected by the VISION survey \citep{Meingast16}. The angular resolution ranges from 2$^{''}$-5$^{''}$ for Spitzer to $\sim$0.7$^{''}$ for VISION, so we can detect more dusty Young Stellar Objects (YSOs) in the same coverage area, down to sizes about 3-7 times smaller where Spitzer excess sources would be blended. \par
In this paper, we present a new catalog of multiple systems in the L1641 and L1647 regions in Orion~A constructed combining the SODA sample \citep{vanTerwisga22} with ancillary ESO-VISTA \citep{Meingast16, Grosschedl19} and Hubble Space Telescope \citep{Kounkel16} observations. The final sample consists only of Class~II YSOs divided into 61 binary systems, 3 triple systems, and one quadruple system, spanning a separation range between 100 and 10$^4$~AU. \par 
The sample selection and the statistical method used to associate the multiple systems are described in Section~\ref{section2}. The results are shown and finally discussed in Sections~\ref{section3} and \ref{section4} respectively. 

\section{Sample characterization} \label{section2}
The focus of this paper is on young dusty sources and their evolution within multiple configurations, so careful selection of an appropriate and unbiased sample is required. \cite{Megeath12} presented a survey of dusty YSOs identified in the Orion~A and B clouds using the IRAC and MIPS instruments on board the \textit{Spitzer Space Telescope}, mapping 9~deg$^2$ in five mid-IR bands from 3–24~$\mu$m, with a resolution of 2$^{''}$-5$^{''}$. The photometric classification of sources is based on IR or mid-IR excess colors, with MIPS and IRAC data merged with the 2MASS point source catalog to generate an eight-band photometric catalog. Among the 298405 point sources identified in the Orion molecular clouds, 3479 sources have been classified as Class II objects (dusty YSOs), with a mid-IR emission above that expected for a reddened photosphere. \cite{Megeath16} slightly updated the sample, including about 10 new dusty sources. Using the Atacama Large Millimeter Array (ALMA), the SODA survey \citep{vanTerwisga22} observed all disks in these previous Spitzer catalogs located in Orion~A below -6$^{\circ}$ degrees in declination. The sample of 873 protoplanetary disks was observed with ALMA at 225~GHz, with a median rms of 1.5~M$_{\oplus}$ (or 0.08~mJy~beam$^{-1}$) - assuming T$_{dust}$=20~K and k$_{\nu}$=2.3~cm$^2$~g$^{-1}$ - and a typical synthesized beam FWHM of $\sim$1.1$^{''}$. Further details can be found in \cite{vanTerwisga22} (Paper~I). \par
Due to the resolution of \textit{Spitzer}, close companions (projected separation $\leq$ 5$^{''}$) are likely to be missing from the infrared catalog. \cite{Grosschedl19} combined archival mid- to far-infrared \citep{Megeath12, Megeath16, Furlan16, Lewis16} and the VISTA telescope data to extend and redefine the catalog of YSOs in Orion A. The updated catalog has the deep seeing-limited resolution of $\sim 0.7^{''}$ and the sensitivity $K_S < 19~mag$ of the ESO-VISTA near-infrared survey of the Orion A cloud (VISION) \citep{Meingast16}, and it covers an extended spatial region of $\sim 950~pc^2$ (or 18.3 deg$^2$). This allows us to construct the largest catalog of dusty YSOs in the Orion~A molecular cloud to date. Compared to 2MASS, the sensitivity of VISTA is 4-5 magnitudes better and its resolution is improved by a factor of about 3. Therefore, the VISION catalog contains more sources in the same coverage area, resulting in an improved YSO classification and a better discrimination of background galaxies or extended nebulous IR emission from YSO candidates. In this way, VISION allows the assembly of the multiplicity sample, while SODA provides the corresponding millimeter data to characterize the structure and evolution of the dust in the circumstellar disk.

\subsection{Catalog Completeness}

The VISION and SODA surveys discuss their catalog completeness with reference to \cite{Megeath16}. The main factor to consider for the completeness of the Spitzer-selected sample is the confusion between the nebular background and the sources, which is estimated from the Route-Median Square Deviations (RMEDSQ) of the pixels surrounding each YSO candidate of the Infra-Red Array Camera (IRAC, \cite{Fazio04}). This provides an estimate of the incompleteness due to the local mid-infrared (MIR) background emission, which varies spatially and increases with stellar density. Thus, \textit{Spitzer} is less sensitive to fainter (or low-mass) stars in regions with non-negligible background emission. This can lead to a bias in our observations: the lower the mass of the star, the fainter and less massive the disk, but with non-negligible dispersion \citep[e.g.,][]{Ansdell16, Pascucci16}. However, this completeness is greatly enhanced in regions where the background emission is faint and the stellar density is low, such as L1641 and L1647. In addition, the VISION survey achieves $\gtrsim$90$\%$ completeness of 20.4, 19.9, and 19.0 mag in J, H, and K$_s$, which allows the observation of a spatially extended sample of YSOs. \par
Given the average distance of Orion~A between 388~pc and 414~pc \citep{Grosschedl18_3D} and the spatial resolution of VISION and SODA of $\sim$0.7$^{''}$ and $\sim$1.1$^{''}$, respectively, we cannot resolve individual components in multiple systems with separation less than 1000~AU. For example, in Figure~\ref{figure:MGM43} (Appendix~\ref{appendixC}) we note that source \textit{Vis2868} seems to be a double system that was not classified as such in either the SODA or VISION catalogs, due to observational limitations. \par
\cite{Kounkel16} presented a near-infrared survey (1.6~$\mu$m) aimed at observing visual multiple systems in Orion at separations between 100 and 1000~AU. Of the 201 protostars and 198 pre-main-sequence stars with disks observed by these authors, they found 29 candidate binary systems and 1 candidate triple system around protostars, and 27 candidate binary systems and 1 candidate triple system around pre-main-sequence stars. In the region of interest, i.e. Orion~A below -6$^{\circ}$ declination, this NIR survey allows us to identify 7 closer binary systems, which are added to our catalog and later included in our analysis (see Table~\ref{table:catalog1}). These double systems have all been observed by SODA, and they are all visible in the 225~GHz continuum ALMA images (see Figures~\ref{figure:MGM950} and \ref{figure:MGM523} in Appendix~\ref{appendixC}), although they were not resolved as a binary system and were therefore classified as single disks in the original VISION and SODA catalogs. \par
\cite{Raghavan10} presented one of the most comprehensive surveys on the multiplicity of solar-type stars, covering stars within 25~parsecs of the Sun. The main conclusion is that the distribution of binary separations ranges from a few AU to over 10$^4$~AU, with a peak around 30-50~AU, suggesting that solar-type stars are commonly found in binary systems with moderate separations, rather than being tightly bound or extremely wide. They also found that $\sim$44$\%$ of solar-type stars have at least one companion, making the occurrence of binary and multiple systems quite common. Therefore, even including the \cite{Kounkel16} visual binaries, our sample is likely to miss half of the binary systems.

\subsection{Multiple systems association}

We take a probabilistic approach to identifying multiple systems, as detailed stellar masses and kinematics cannot be obtained for most of the sample. The probability that two objects are physically associated, indicating a true multiple system, decreases with the square of the distance between them. \par
To investigate the stellar multiplicity in our sample, we employ the iterative inside-out search method presented by \cite{Tobin_2022}. Starting from an initial search radius of 1000~AU, which by observational limitations represents the minimum observable separation in our sample, we iteratively search for companions with an increasing separation up to 10$^4$~AU. Beyond this upper limit, we consider the probability of finding a physically bound system negligible \citep{Tobin_2022}. Therefore, we search in the VISION catalog \citep{VisionCatalogue19} for the nearest Class II neighbor to each Class II source identified by \textit{Spitzer} in regions L1641 and L1647 using the k-nearest neighbors (KNN) algorithm. \par
When two sources are associated by the algorithm, they are considered as a candidate binary system. Their individual catalog entries are replaced with a new single one, whose coordinates corresponds to the geometric midpoint of the newly formed system without any weighting. These multiple systems can be further associated with other individual sources or multiple systems. By removing the individual entries, we ensure that each component is associated with only one multiple system, avoiding multiple assignments. Since the only imposed condition is on distance, there is no upper limit on the number of potential associations. \par
It is important to note that the algorithm only checks the separation between the two objects forming the multiple system at a given step, but not the distance between the individual objects within the multiple systems. This means that if, for example, we consider a triple system, the distance between the three individual objects can be greater than the chosen limit, as long as the separation between the binary system and the single source forming the triple is less than 10$^4$~AU. \par
This method is not limited to a specific observation wavelength and offers simplicity in replication, but in order to distinguish the bound pairs from chance alignments, it is necessary to find a way to assign a measure to each system to distinguish between the two cases. We choose to consider the probability of random alignments by considering the local surface density. The procedure is described in Section~\ref{section2.3}.

\subsection{Multiple system selection} \label{section2.3}

Chance alignments with other YSOs can introduce contamination into the observed multiple systems within the sample. This likelihood of contamination generally rises with larger separations and higher local surface densities of YSOs. To address this issue, it is necessary to assess the local surface density around each target in the sample and the probability of detecting an unassociated source. \par
The local surface density $\Sigma_{\small{YSO}}$ is computed employing the KNN algorithm. This involves dividing the arbitrary number of neighbors by the area of the circle with a radius equal to the distance (in parsecs) between the central object and the farthest selected object. The resulting value provides an estimate of the expected number of YSOs within a given area surrounding the central target. Following \cite{Megeath16}, we select 10 neighbors, obtaining:
\begin{equation}
    \Sigma_{\small{YSO}} = \frac{10}{\pi R_{10th}^2}.
\end{equation}
In principle, this approach can cause problems along the edges of the entire sample region. For objects along the periphery, the i-th neighbor may not represent the true i-th companion, since elements outside the region may not be considered. As a result, the local density may be underestimated. However, our SODA sample is mainly located in L1641 and L1647 (with a FoV of $\sim$50$^{''}$), while VISION, i.e., the catalog we use to identify multiple systems, has a larger spatial footprint ($\sim$19.8~deg$^2$), and it is therefore complete when searching for our nearest companions. \par
In our framework, we make use of the following statements:
\begin{itemize}
    \item [D:] The object $i$ is within the area $A$;
    \item [C:] The object $i$ is the companion (secondary) of the object $k$;
    \item [I:] The primary object $j$ is detected (background information).
\end{itemize}
\par
The probability of whether a detected source is a companion or not is determined by Bayes' Theorem
\begin{equation}
    P~(C|D,I) = \frac{P~(D|C,I)~P~(C)}{P~(D)}.
\end{equation}
\par
In our notation, $P(D|C,I)$ is the probability of detecting a companion if this is actually present. We assume the NIR catalog is complete for Class II sources and thus $P(D|C,I)$~=~1. \par
$P(D|I)$ is the combined probability of detecting a source, whether it is a companion or an unassociated source, within a given research radius $d$. To derive these terms, we followed \cite{Tobin_2022}. The detection probability is modeled as a Poisson distribution
\begin{equation}
    P~(k) = \frac{\lambda^k e^{-\lambda}}{k!},
\end{equation}
where $\lambda = \sum_{YSO} \pi d^2$, $d$ the search radius employed in the density calculation and $k=10$ is the number of expected YSOs within this radius. Therefore, the probability of detecting at least one unassociated source is 
\begin{equation}
    P~(unassociated~YSO) = 1 - P(k=0) = 1 - e^{\sum_{YSO} \pi d^2},
\end{equation}
the overall probability of detecting an object, regardless of its state of association, is
\begin{equation} 
    P~(D|I) = P(~C|I) + (1 - e^{-\Sigma_{YSO}\pi d^2}) (1 - P~(C|I)),
\end{equation}
\par
and the probability for this object of being the secondary of a binary system depends only on separation and on the local surface density of YSOs
\begin{equation}
     P~(C|I) = e^{-\Sigma \pi d^2}.
\end{equation}
\par
We computed the probability of being a multiple system for all the possible multiples, including in our final catalog only the ones that pass the threshold value of 0.9. In the absence of any other information, such as relative velocity or similar, and thus unable to test the robustness of this approach, the best choice is to consider a high threshold (P > 0.9) which minimizes false positives while maintaining a high true positive rate.

\section{Results} \label{section3}

\begin{figure*}
\centering
\subfigure{
   \includegraphics[width=2\columnwidth]{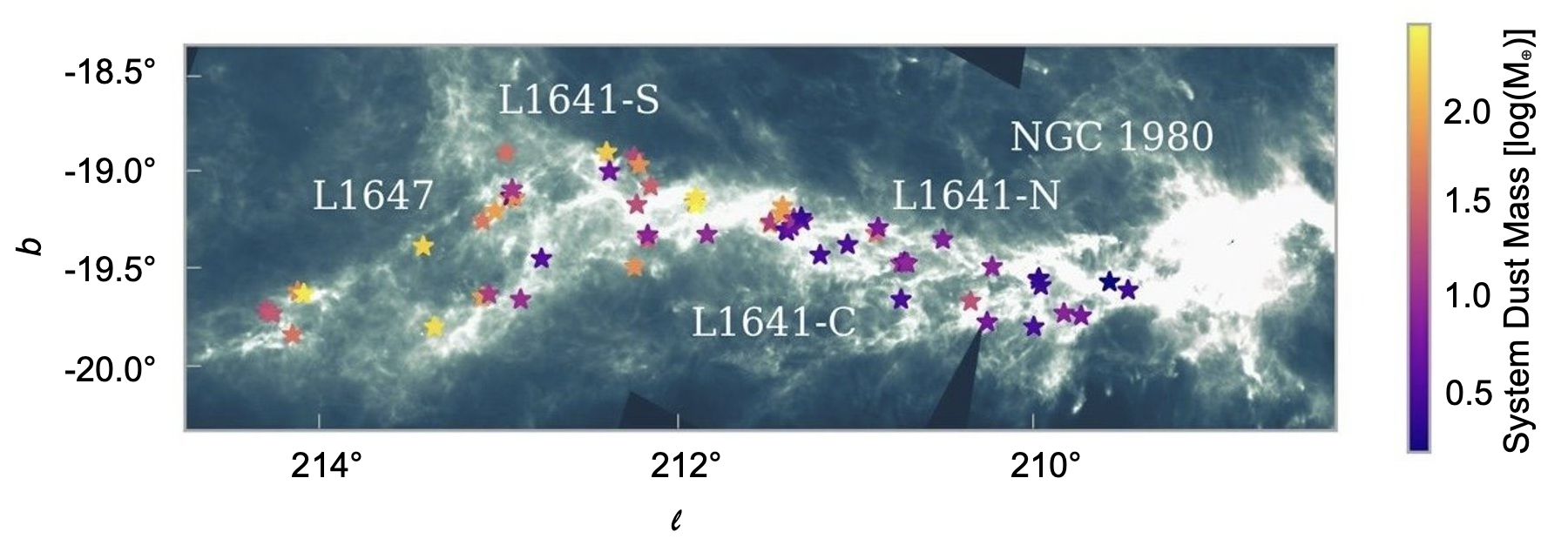}
     \label{figure:Herschel1}
}
\subfigure{
   \includegraphics[width=2\columnwidth]{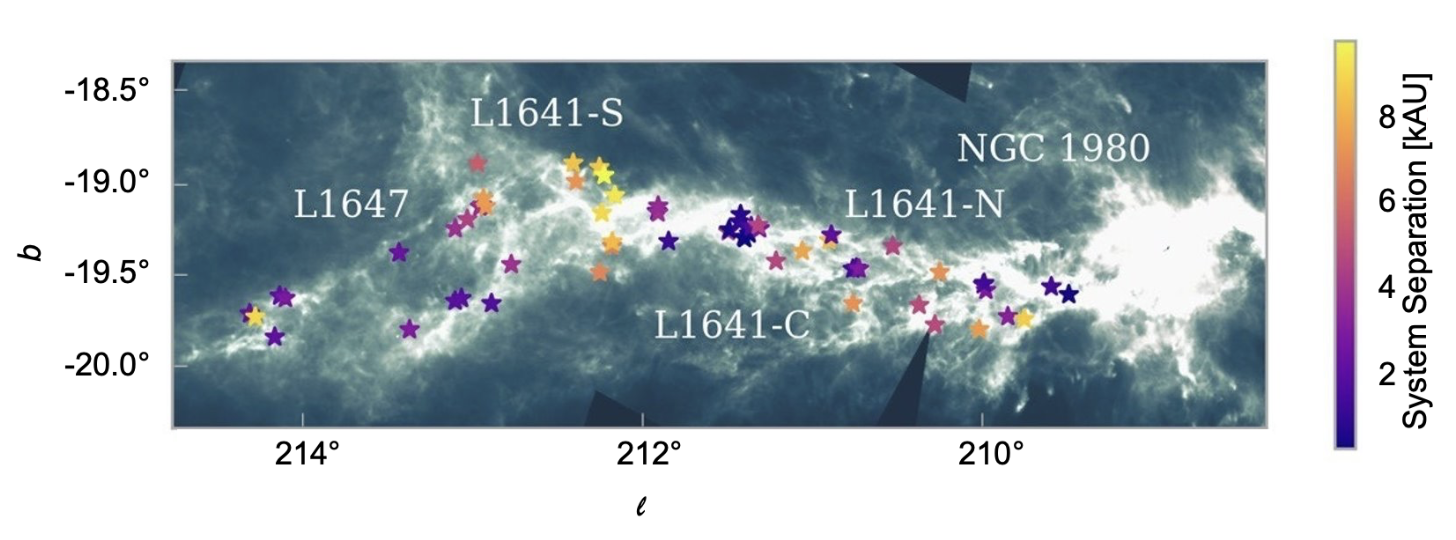}
   \label{figure:Herschel2}
}
     \caption{\textit{Top:} Spatial distribution of binary systems in Orion~A, color-coded according to the system disk dust mass. \textit{Bottom:} Spatial distribution of binary systems in Orion~A, color-coded according to the system separation. \\
     Stars mark the position of the binary systems. \textit{Herschel} SPIRE observations at 250, 300, and 500 $\mu$m \citep{Soler19} form the background.}
\end{figure*}

Our final catalog counts 61 binary systems, 3 triple systems, and 1 quadruple system, located in the regions L1641 and L1647 of Orion~A. Figure~\ref{figure:Herschel1} shows the sample distribution of our binary systems in Orion~A color-coded according to the system mean disk dust mass, whereas Figure \ref{figure:Herschel2} shows the distribution of the same sample color-coded according to the system separation. It appears that the lower mass binaries are mainly located at lower longitudes, while the higher mass ones are at higher longitudes. On the other hand, systems with separations smaller than about 3~kAU are evenly distributed along the Orion~A filament, while those with the largest separations are mainly located in L1641-S and L1641-N. \cite{VanTerwisga23} have demonstrated that L1641-S and L1641-N contain the most irradiated disks in Orion~A. Therefore, our separation distribution could be a result of the external far-ultraviolet (FUV) radiation: the presence of massive stars could influence the system formation process or its very early evolution. \par
In this paper we focus primarily on disk dust mass in multiple systems. Further comments and analysis on system separation are presented in Appendix~\ref{appendixA}. 

\begin{figure*}
    \centering
    \includegraphics[width=15cm]{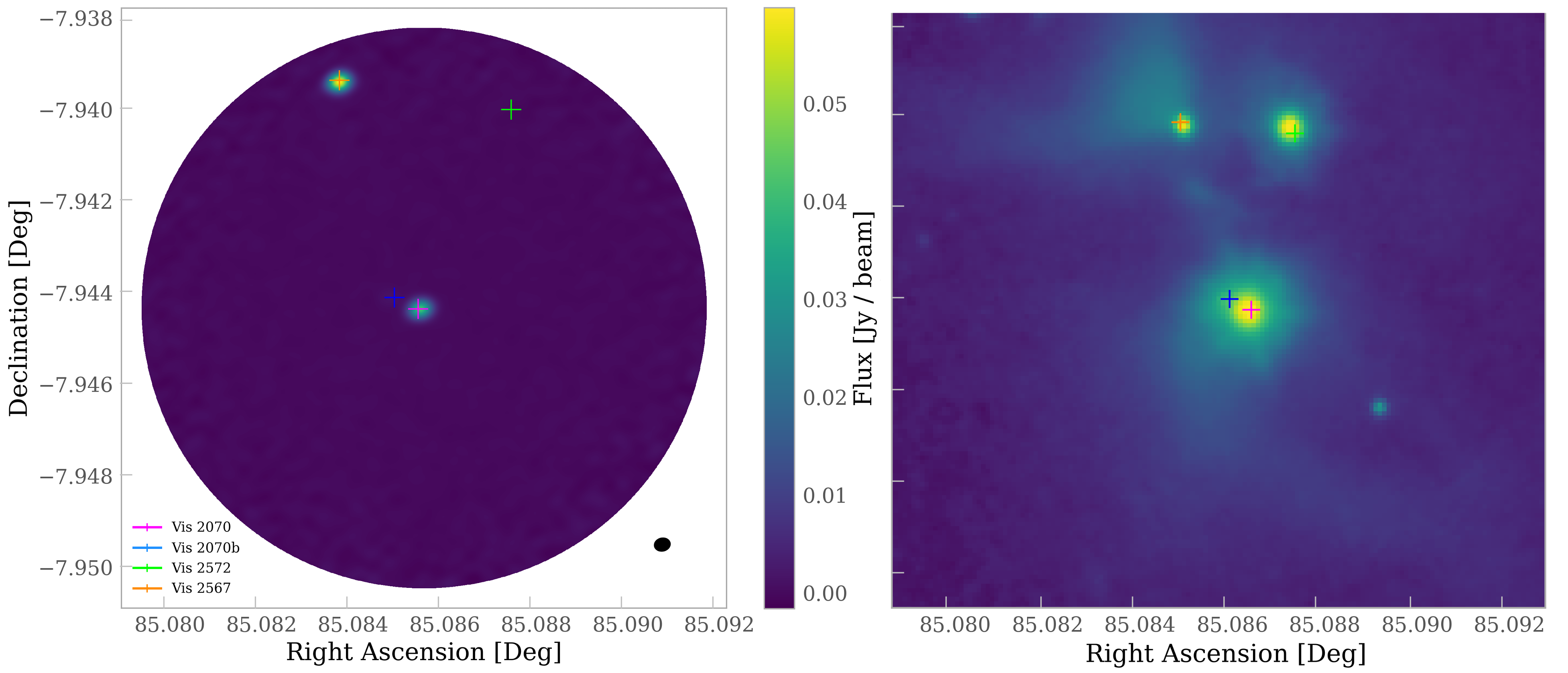}
    \caption{\textit{Left:} Continuum ALMA image at 225~GHz from the SODA survey \citep{van_Terwisga_2022}. The crosses mark the positions of the four sources in the quadruple system, identified by the source index in the VISION~III catalog \citep{Grossschedl20}. The central pair [Vis~2570] - [Vis~2570b] has only been resolved by \cite{Kounkel16}. The beam is shown in black in the right-hand. \textit{Right:} Image drawn from the ESO-VISTA near-infrared survey of the Orion A cloud of the same system as in the left-panel (\cite{Meingast16}).}
    \label{figure:quadruple}
\end{figure*}

The quadruple configuration is rarely observed in the other nearby Star-Forming Regions (SFRs). Figure~\ref{figure:quadruple} shows the only quadruple system in our analysis, consisting of 4 young stars (ages $\sim$1-3~Myr), each of which is surrounded by a disk 
detected in the mid- to far-infrared by the VISION survey, but not always detected and resolved at 225~GHz with ALMA \citep{van_Terwisga_2022}. Such a configuration provides a rare opportunity to better constrain the possible effects of gravitational interactions between disks within the same system. \par
For reference, some of the assembled multiple systems are shown in appendix~\ref{appendixC}.

\subsection{Disk dust mass}
As discussed in section~\ref{section2}, the sample was assembled considering the VISION data. This choice allows us to construct a more complete sample down to smaller separations. Nevertheless, more compact systems (separation $\leq$1000~AU) cannot be resolved in our sample and are usually classified as single objects. To solve this problem, our analyses include 7 new closer binaries, with separations between 100 and 1000~AU, resolved by \cite{Kounkel16}. All of these sources are visible, but not always resolved, in the continuum ALMA images at 225~GHz. We estimate the disk masses based on their ALMA continuum using the same procedure presented in paper~I \citep{van_Terwisga_2022}. \par
We assume that the continuum millimeter flux is optically thin. Knowing the distance $d$ and the opacity $k_{\nu}$, the millimeter continuum flux $F_{\nu}$ can be related to the mass of the dust emitting the radiation by the equation:
\begin{equation} \label{equation:3.1}
    M_{dust} = \frac{F_{\nu} d^2}{k_{\nu} B_{\nu}(T_{dust,eff})}.
\end{equation} 
For opacity and dust effective temperature we consider the constant values of $k_{\nu}$=2.3~$cm^2~g^{-1}$ and $T_{dust,eff}$=20~$K$, respectively, chosen in paper~I. We also rely on paper~I for the disk distances, and the resulting range, between 380 and 455 parsecs, has also been adopted in this work. \par
Table~\ref{table:catalog1} lists average distances, projected separations and mean dust masses of our multiple systems. It should be noted, however, that the assumptions of optically thin emission and constant opacity introduce an unquantifiable source of uncertainty. There is increasing evidence that a fraction of the millimeter part of the disk structure may be optically thick \citep[e.g.,][]{Xin23, Zhu19}, and that the assumed opacity value of millimeter grains along the disk structure may not always be valid or constant \citep[e.g.,][]{Krapp22, Macias21, Zhu19}. As a result, the disk dust masses may be underestimated. However, if the disk emission is (mostly) optically thick at these wavelengths, the millimeter continuum flux is instead tracing the radial extent of the dust in these sources.

\section{Discussion} \label{section4}
In recent years, several surveys have been conducted in nearby SFRs to characterize the structure and better understand the evolution of protoplanetary disks through all phases of planetary formation \citep[e.g.,][]{Tobin_2022, van_Terwisga_2022, Zurlo21, Manara19}. The results shed new light on key aspects such as the spatial distribution of gas and dust mass, the evolution and dynamics of the gas and dust components within the disk, or the links between the disk structure, the host star, and the planet system. Nevertheless, the available catalogs are not large enough and do not span a sufficiently large range in separations to study how multiplicity affects disk survival. \par
In this paper we present the largest catalog of Class II YSOs multiple systems in Orion~A, about 94$\%$ of which are binary systems. Our analysis focuses on the disk dust masses, since this is undoubtedly one of the properties of crucial interest for disks, providing elementary constraints on the future content of planetary systems. 
Recent observations show that physical interactions within multiple systems \citep[e.g.,][]{Zagaria21, Manara19, Cox17} and (F)UV radiation \citep[e.g.,][]{VanTerwisga23, Winter22} significantly affect the disk masses. \par
In the following sections, we prove that there is a projected separation beyond which the millimeter flux (and hence the dust mass) and the separation are no longer correlated, which implies that at large separations the evolution of disks is independent of the presence of a companion. We obtain this conclusion by combining observations of protoplanetary disks in multiple systems in Orion, Lupus, and Taurus. We also show that the mass distribution of the disks in binary systems is the same in regions of similar age and that dust evolution can therefore be explained as a function of age. Finally, by focusing on the effects of stellar multiplicity on the individual components of binaries, we show that it is statistically impossible to distinguish between disks around single stars and disks around stars in multiple configurations in the SODA sample.

\subsection{Flux-Separation correlation} \label{section:fluxseparation}
In the past decade, several (sub-)millimeter surveys of protoplanetary disks in multiple systems have focused on the relation between the radii and masses and their projected separations \citep[e.g.,][]{Zagaria21, Manara19, Ansdell16}. Among these studies, \cite{Harris_2012} presented a high angular resolution millimeter-wave dust continuum imaging survey of circumstellar material associated with the individual components of 23 multiple star systems in the Taurus–Auriga young cluster, proving that the millimeter flux of the combined system positively relates to the projected separation of the stellar companion. Subsequently, \cite{Zagaria21} proves that this same relation holds for a larger sample of Taurus disks, and if $\rho$ Ophiuchus and Lupus samples are considered as well. \par
\begin{figure}
    \centering
    \includegraphics[width=\columnwidth]{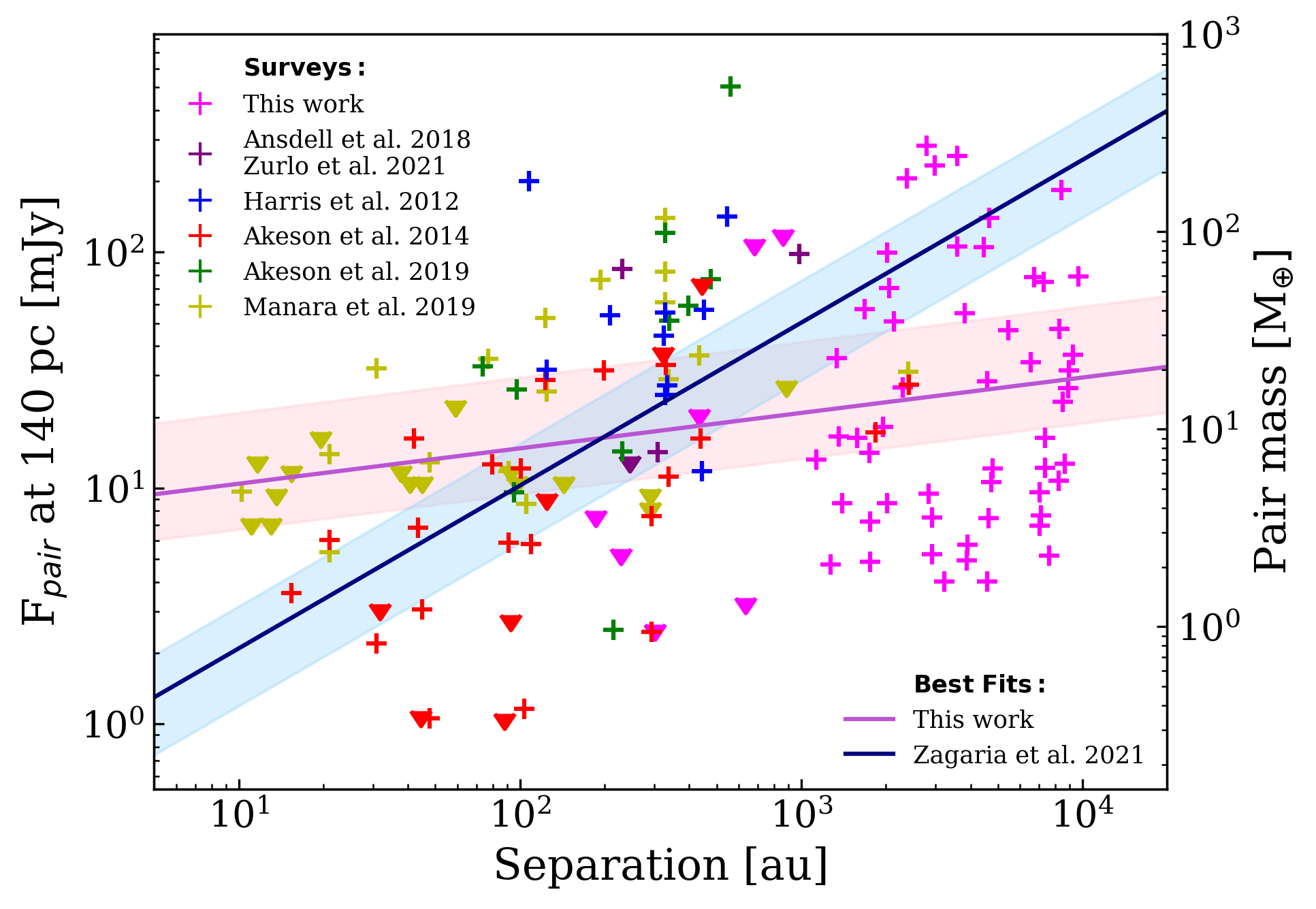}
    \caption{Pair millimeter flux (re-scaled to a distance of $d=140~pc$) and dust mass of binary systems as a function of their projected separation, located in Orion~A, Lupus \citep{Zurlo21, Ansdell18}, and Taurus \citep{Akeson19, Manara19, Akeson14, Harris_2012}. Crosses represent detections and triangles represent non-detections: according to \cite{Akeson19}, we consider a binary pair to be detected only if both disk components were detected. In blue is the linear regression (slope $m=0.69^{+0.13}_{-0.12}$, intercept $q=-0.45^{+0.31}_{-0.33}$) from \cite{Zagaria21}, and in purple is the linear regression (slope $m=0.15^{+0.11}_{-0.12}$, intercept $q=0.92^{+0.34}_{-0.32}$) we get considering also our binary sample.}
    \label{figure:fluxsep}
\end{figure}

Comparing the properties of all the disks in Orion with those in Taurus and Lupus, we find essentially no difference in disk mass distribution or abundance (e.g., \cite{van_Terwisga_2022}). Furthermore, all these regions are of similar age and have similar underlying stellar mass distributions (\cite{DaRio}). Thus, while the Orion sample is restricted to wide binaries, and the Lupus and Taurus samples to close binaries, the evidence suggests that the underlying populations have similar properties. It is therefore worth investigating what happens when the Orion region is included. In Figure~\ref{figure:fluxsep}, we have reproduced the same plot as \citeauthor{Zagaria21}, considering the multiple sample in Lupus \citep{Zurlo21, Ansdell18}, Taurus \citep{Akeson19, Manara19, Akeson14, Harris_2012}, and the regions L1641 and L1647 in Orion~A. As a guide, we show the linear regression from \cite{Zagaria21} - performed using the hierarchical Bayes model \texttt{Linmix}~\footnote{https://github.com/jmeyers314/linmix} \citep{Kelly_2007} -  that also includes the $\rho$ Ophiuchus region (see Figures 1 and 3 of the paper), and the linear regression we obtain considering also our wider binary sample in Orion~A. It is readily apparent that the latter trend deviates from the previous one, and that a global fit to all data is almost flat. 
Indeed, looking at Figure~\ref{figure:fluxsep}, it appears that at higher separations the data points are more evenly distributed across the graph, so that the previously observed flux-separation correlation may be lost. \par
\begin{figure}
    \centering
    \includegraphics[width=\columnwidth]{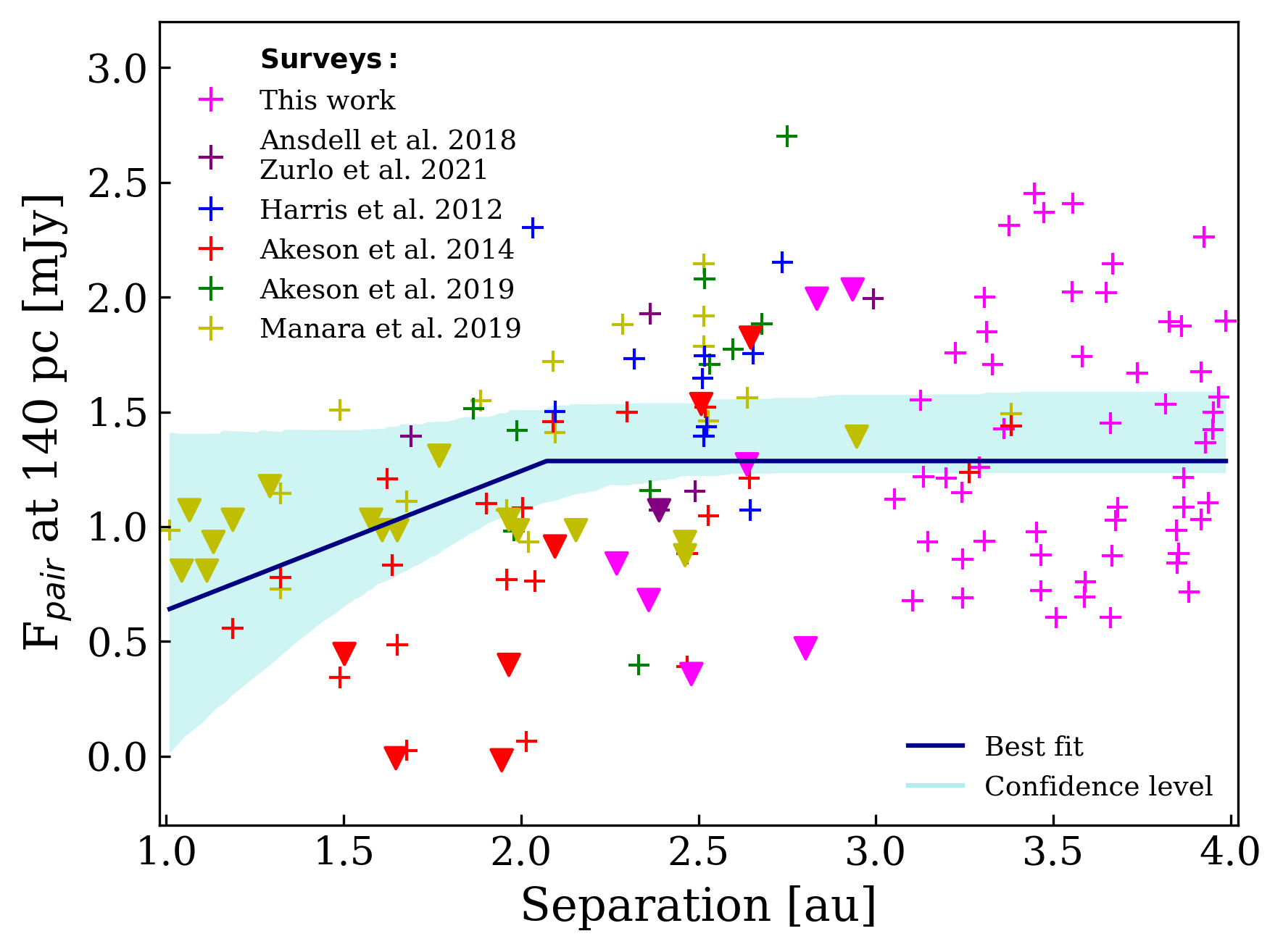}
    \caption{Regression best-fit of the logarithm of flux over the logarithm of separation. Free parameters: m (slope),  q (intercept), and x$_{edge}$ (i.e., separation threshold).\\ Crosses indicate detected systems, while triangles indicate upper limits. The binary systems are color-coded as in Figure~\ref{figure:fluxsep}.}
    \label{figure:linearconst}
\end{figure}

To quantify this behavior, we perform an additional fit to the combined observations, including a separation threshold beyond which we assume a constant system dust mass. The fit is shown in Figure~\ref{figure:linearconst}. Table~\ref{table:bestfits} lists the best-fitting parameters of our two fits, and the respective corner plots are shown in Appendix~\ref{AppendixCorn}. 

\begin{table}[hbt!]
\caption{Regression parameters for our fits in Fig.s~\ref{figure:fluxsep}-\ref{figure:linearconst}.} 
\label{table:bestfits}
\centering  
\begin{tabular}{l l l l l} 
\hline\hline  
 & ~~~~~m & ~~~~~q & ~~~~~~x$_{edge}$  & ~~~$log$~Z \\
 & \textit{~~~Slope} & \textit{~Intercept} & \textit{Separation} &\textit{Marginal} \\
 & [$mJy/au$] & ~[$mJy$] & \textit{~threshold} &\textit{likelihood} \\
\hline\hline 
    & & & \\
   \small{Linear Fit} & \small{0.15$^{+0.11}_{-0.12}$} & \small{0.92$^{+0.34}_{-0.32}$} & ~~~~~~- & \small{-20.831$^{+0.101}_{-0.101}$} \\ 
   ~~\small{(\textit{Fig.~\ref{figure:fluxsep}})} & & & \\
   \small{Broken Fit} & \small{0.64$^{+0.65}_{-0.45}$} & \small{0.06$^{+0.86}_{-0.90}$} & \small{2.10$^{+0.58}_{-0.64}$} & \small{-18.333$^{+0.109}_{-0.109}$} \\
   ~~\small{(\textit{Fig.~\ref{figure:linearconst}})} & & & \\
\hline
\end{tabular}
\end{table}

The plots are obtained using the \texttt{UltraNest}\footnote{https://johannesbuchner.github.io/UltraNest/readme.html} package by \cite{Buchner16}, which allows the fitting and comparison of complex models and implements a Monte Carlo technique called Nested Sampling. Specifically for our case upper limits are treated as constraints on the parameter space. The UltraNest algorithm iteratively samples from a series of nested probability contours, gradually exploring the imposed parameter space. During each iteration, UltraNest calculates the marginal likelihood $Z$ of the sampled parameter values based on the data and the imposed constraints, allowing the posterior distribution to be effectively explored. It is then possible to determine the relative predictive power of different models using Bayes Factors ($BF$). \par
The linear regression in Figure~\ref{figure:fluxsep} yields $log~Z=-20.831\pm0.101$, while the regression fit in Figure~\ref{figure:linearconst} yields $log~Z=-18.333\pm0.109$. Therefore, $BF=exp(log~Z_2 - log~Z_1)~>~1$, confirming that the second model is substantially favored (see \citeauthor{Jeffreys98}~1998): we identify a threshold separation x$_{edge}$~=~127~$\pm$~4~AU beyond which the previously observed positive correlation between pair millimeter flux and projected separation is lost. 

Recent theoretical models and simulations show that multiple systems are formed mainly through three channels: turbulent fragmentation of the molecular cloud \citep[e.g.,][]{Kawasaki23}, thermal fragmentation of strongly perturbed, rotating, and infalling core \citep[e.g.,][]{Boss13, Boss14}, and fragmentation of a gravitationally unstable circumstellar disk \citep[e.g.,][]{Longarini23, Stamatellos09}. The first two scenarios will lead to multiples that are initially separated by a few hundred to a few thousand AU, while the last one leads to the formation of companions with separation of $\leq$100~AU \citep[e.g.,][]{Tobin13, Takakuwa12}. \par
From this perspective, our results suggest that the systems before and after the threshold are the outcome of different formation processes. In particular, before the separation threshold we can consider the multiple system to be the result of the fragmentation of a gravitationally unstable circumstellar disk, while after this threshold, it is likely the result of the thermal fragmentation of an infalling core and the turbulent fragmentation of a molecular cloud. \par
In the following sections, we explore a number of alternative possible causes that could account for the different observed flux-separation correlations. First, we compare the mass distributions in the nearby SFRs aforementioned with those in Orion~A (see Section~\ref{section:nearbySFRs}), then we analyze what happens when we consider the components of the binary systems separately (see Section~\ref{section:Single&Binary}). However, the results obtained cannot explain the behavior highlighted in Figure~\ref{figure:fluxsep}. \par 
The broken flux-separation correlation may imply that the mechanisms behind the formation of multiple systems at fixed ages are affected by external environmental influences, with dramatic effects on planet formation. \cite{VanTerwisga23} showed that FUV radiation from the A0 and B massive stars, mainly located in the Orion Nebula Cluster (northern part of Orion~A), significantly reduces the mass of the disks, with the disks losing a factor of $\sim$2 in mass over two orders of magnitude in FUV field strength. However, we do not observe a significant mass-separation trend in the SODA sources alone, and the similarity of the cumulative disk mass distributions suggests that the FUV radiation does not dominate the results. \par
The effect of FUV on the multiplicity properties of YSOs has not been well studied, and we do not have the data to test this in more detail. In addition, the results of our analysis are inherently limited by the observed systems: the observational resolution allows us to observe only systems with large separations. A closer comparison between systems of the same age and separation to the Taurus and Lupus data is possible with ad hoc spectroscopic surveys or direct imaging.

\subsection{Mass distribution of multiple systems in nearby SFRs} \label{section:nearbySFRs}

The result presented in the previous section was obtained by considering three different SFRs. Therefore, the new trend observed could be due to environmental differences in different regions. To clarify this, in this section, we compare the disk mass distributions of binaries in the 4 nearby SFRs Lupus, Taurus, Orion~A, and Upper Scorpius. \par
We follow other studies \citep[e.g.,][]{Zurlo21, Testi22, vanTerwisga22, Grant21, Cazzoletti19} in using the Kaplan-Meier estimator to infer the distribution of disk masses. This work uses the implementation of this estimator in the \texttt{lifelines} package \citep{cameron_davidson_pilon_2020_3629409}. This non-parametric tool for assessing censored random variables, relying on the assumption that the censoring is unrelated to the variable of interest, is robust in describing these distributions and facilitates their comparison. \par
\begin{figure}
\centering
   \includegraphics[width=\columnwidth]{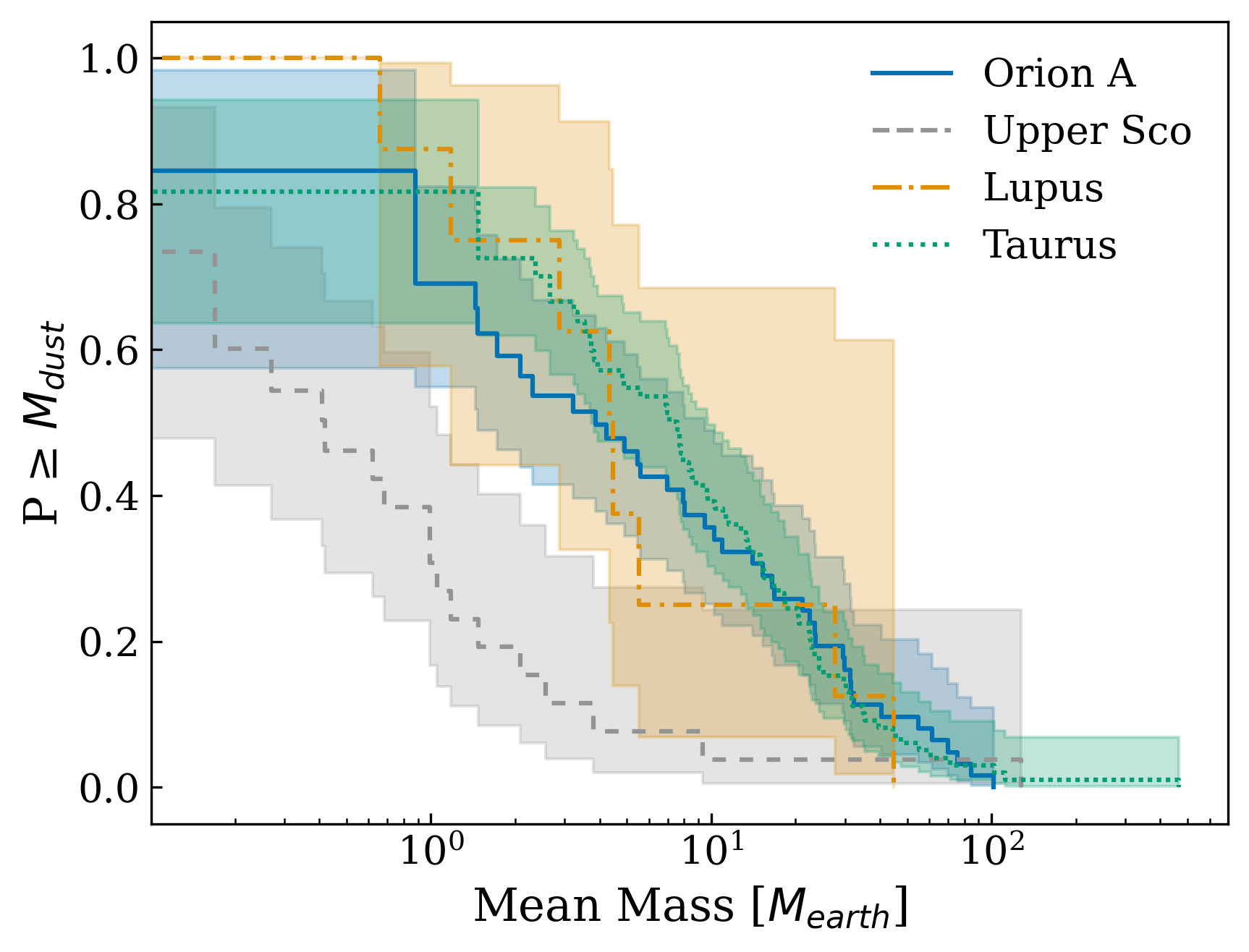}
     \caption{Dust mass distributions inferred with the Kaplan-Meier estimator for the binary systems located in Orion~A, Lupus, Taurus, and Upper Scorpius. For the later 3 regions, we have considered the available data from the relevant literature \citep{Zurlo21, Ansdell18, Akeson19, Manara19, Akeson14, Harris_2012, Barenfeld_2019}.}
     \label{figure:literature}
\end{figure}

Figure~\ref{figure:literature} shows the inferred cumulative mass distribution for the approximately same-age Lupus, Taurus, and Orion A (age$\simeq$1-3~Myr), and the older Upper Scorpius (age$\simeq$5-11~Myr). Looking at regions of similar and different ages allows us to address the dependence between the region's age and the evolution of the disks within it. 

The cumulative trends in Figure~\ref{figure:literature} show a consistent disk evolution across the whole mass range for the younger regions. To quantify this similarity, we use the log-rank correlation test \citep{Feigelson85}, which states that a significant difference between samples can only be detected when the p-value is $p \leq 0.05$. Note that since \texttt{lifelines} only supports right censored data for this test, we subtract our data from a constant before running the test. For our comparison, we find a p-value $\geq 0.05$ for each pair combination, indicating that the cumulative distributions for Lupus, Taurus, and Orion~A are not statistically distinguishable. On the contrary, the log-rank test confirms that the distribution of Upper Scorpius is statistically different from the distributions observed in the younger SFRs: the disks in the older region have lower dust masses, confirming that dust evolution is even more advanced in this region. This result is consistent with what has already been observed: recent surveys prove that if disks are optically thin at millimeter wavelengths, they typically hold just a few Earth masses of dust available for potential planet formation within the first 1-3~Myr \citep[e.g.,][]{Ansdell18, Andrews13}, while in later stages the dust evolution proceeds even further, resulting in even lower median disk \citep[e.g.,][]{Barenfeld16}.

\subsection{Mass distribution in Orion A: separate system components} \label{section:Single&Binary}

The results shown in Sections~\ref{section:fluxseparation} and \ref{section:nearbySFRs} are obtained by considering the binary systems as a whole, i.e., the analyses consider the sum of the fluxes and masses of the two components of the systems. In this section, we look at how stellar multiplicity affects the individual components of binary systems. Our multiple sample results in mainly binary systems. For each of these systems the primary disk is selected as the brightest: considering the J magnitude in VISION III -- or the H and K magnitudes if the J magnitude is not available -- the disk with the lowest magnitude is selected as the primary. \par
\begin{figure}
\centering
   \includegraphics[width=\columnwidth]{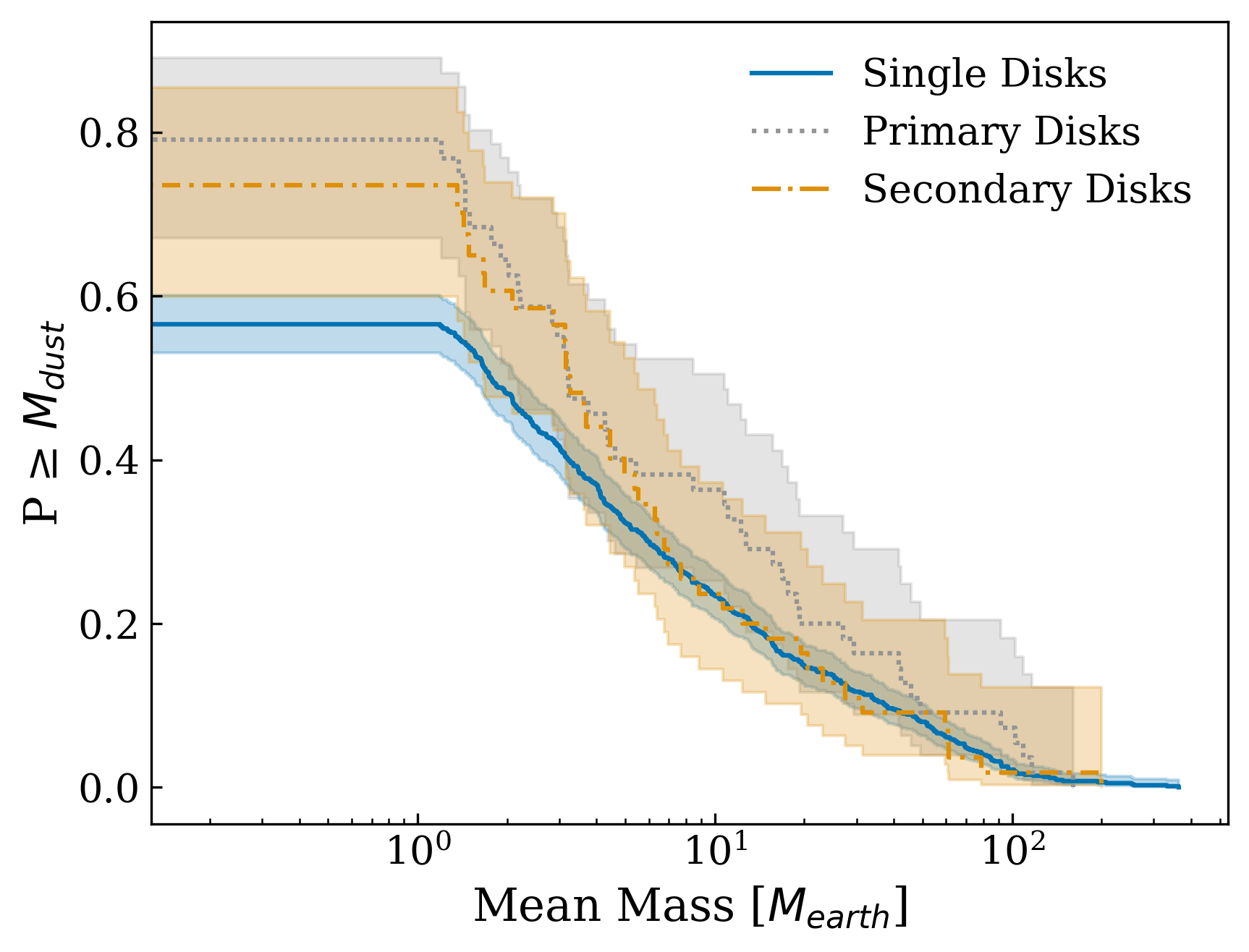}
     \caption{System mean dust mass distributions inferred with the Kaplan-Meier estimator for primaries, secondaries and single Class~II disks in L1641 and L1647, Orion~A.}
     \label{figure:sepbinaries}
\end{figure}
\begin{figure}
\centering
   \includegraphics[width=\columnwidth]{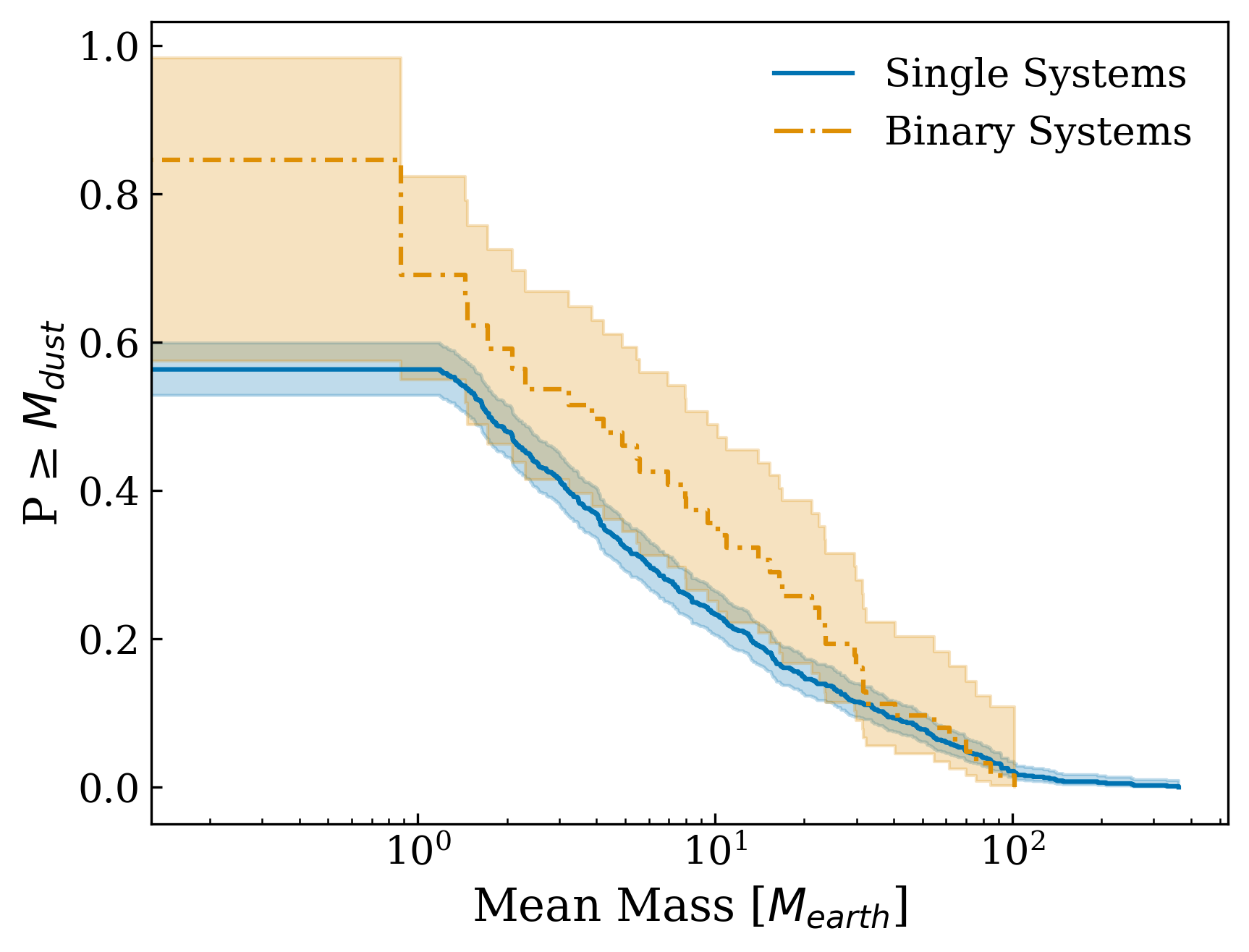}
     \caption{System mean dust mass distributions inferred with the Kaplan-Meier estimator for binary and single systems in L1641 and L1647, Orion~A.}
     \label{figure:cumulativemass}
\end{figure}
\begin{figure}
\centering
   \includegraphics[width=\columnwidth]{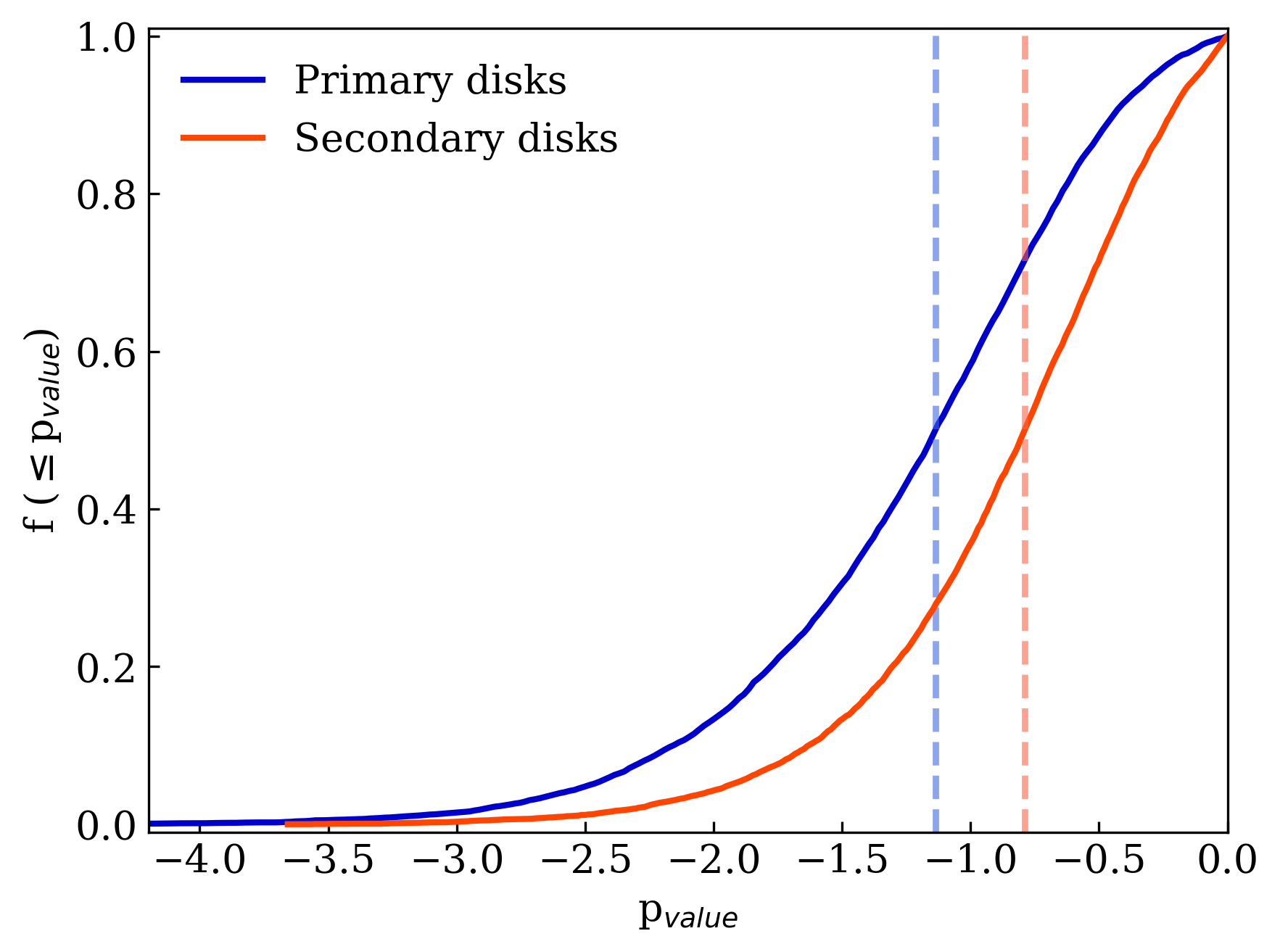}
     \caption{Cumulative distributions of p-values for the comparison between the mass distribution of primary and secondary disks and an equal-size sample of randomly selected single disks with similar J-band magnitude distribution. The distributions are made by repeating the test 10$^4$ times. The median p-values for the primary sample of 0.071 and for the secondary sample of 0.166 are shown for reference.}
     \label{figure:pvaluetest}
\end{figure}

Figure~\ref{figure:sepbinaries} shows the cumulative distributions of the primary, secondary, and single disk dust masses. As a comparison, Figure~\ref{figure:cumulativemass} shows the cumulative mass distributions for the binary systems as a whole, i.e. we refer to the unweighted average mass of the dust in the two disks, and the single disk systems. The latter comparison suggests that the most massive disks surround single stars and that the derived distributions for binary and single systems are different. Although the two distributions are compatible within the 1~$\sigma$ confidence intervals, they differ both in the region $\leq$~10$^2$~M$_{\oplus}$, where the binary systems are more likely to have a higher mean mass, and in the region $\geq$~10$^2$~M$_{\oplus}$, where the distribution of the pairs ends. The log-rank test yields a p-value of $\sim$~0.023, which suggests a (weak but significant) difference in the underlying distributions, and thus the disk evolution depends on the presence of a companion and the resulting physical interactions with it. For Figure~\ref{figure:sepbinaries}, on the other hand, the log-rank test proves that only the primary and the single disk distributions are statistically distinguishable ($p \sim$~0.036). Given the relatively low significance of the log-rank tests and the limited sample sizes, we prefer a more robust statistical comparison between the mass distributions in Figure~\ref{figure:sepbinaries}. We expect primary stars to be more massive on average than secondary stars, by definition. Spectroscopy-based stellar mass estimates are not available yet for the majority of the multiple system candidates: from the Sloan Digital Sky Survey (SDSS) APOGEE INfrared Spectroscopy of Young Nebulous Clusters program (IN-SYNC) of the Orion~A molecular cloud \citep{DaRio} we obtained information for both components for only 6 pairs and information for only one component for 19 systems. These stellar parameters are listed in Table~\ref{table:stellarmass}, and according to this, we computed a mean age of 1.2~Myr for our binary sample. To ensure a more robust comparison, we adopted a bootstrap testing approach to validate the results. Specifically, we selected identical-size subsets of disks from the reference single-disk sample with similar J-band magnitude distribution of primary and secondary disks. For each resampled subset, we compared the dust mass distributions with the original distributions using the log-rank test. We repeated this procedure 10$^4$ times to generate a bootstrap distribution of p-values. The results are shown in Figure~\ref{figure:pvaluetest}. The median p-value for the primary and secondary disks is about 0.07 and 0.16, respectively, proving that the simulated and original mass distributions are statistically indistinguishable. This means that in our sample there are no significant differences between the disks in isolated and multiple systems for the separation range considered. We expect such differences for more compact multiple YSOs, but due to the limitations of our observations, we cannot reach these lowest separation ranges.

\section{Conclusions} \label{section5}

We have characterized the stellar multiplicity of regions L1641 and L1647 in Orion~A. The resulting catalog contains all Class II disks divided into 61 binary systems, 3 triple systems, and one quadruple system. The separation range covered is from 100 to 10$^4$~AU. This is the largest sample assembled in the literature to date. We characterize the dust mass distributions for single disk and multiple systems. Our statistical analyses allow us to quantify the differences between these two samples and also to compare the disk evolution we obtain for the multiple sample with the disk evolution obtained for the multiple samples in other nearby SFRs. \par 
Our main findings are as follows:
\begin{enumerate}
    \item Recent studies \citep[e.g.,][]{Harris_2012, Zagaria21} show a positive correlation between the millimeter flux (and hence the dust mass) and the projected separation of binary systems in the nearby SFRs Lupus and Taurus. Our data in Orion A alone, however, suggest a weaker correlation and are consistent with no correlation within the confidence interval: although the regression fit is inconclusive since we cannot exclude a slope equal to zero, the positive correlation within the 2 sigma range physically suggests that as the separation increases, the physical interaction between the system components becomes negligible;
    \item When considering the joint data of Lupus, Taurus, and Orion~A, the positive flux-separation correlation is lost at a threshold separation of about 130~AU; 
    \item Recent simulations and theoretical models have confirmed that different star formation processes lead to multiple systems with different separations, and from this perspective, our result suggests that before and after the threshold of 130~AU we can refer to fragmentation of a gravitationally unstable circumstellar disk, and thermal or turbulent fragmentation, respectively;
    \item The cumulative distributions for binaries in the younger regions of Orion~A, Lupus, Taurus, and the older Upper Scorpius prove that the disk evolution within regions of the same age is consistent across the mass range: the log-rank test confirms that the only statistically distinguishable distribution is Upper Scorpius, the oldest star-forming region, strengthening the idea that there is a universal initial mass distribution for disks;
    \item The comparison between the mass distributions of the primary and secondary disks with an equal-sized sample of randomly selected single disks with similar J-band magnitude distributions confirms that it is not possible to statistically distinguish the dust evolution of the disk in single and multiple for the separations considered in this study (10$^3$ - 10$^4$~AU); 
    \item We found a unique system: a quadruple system consisting of 4 stars, all surrounded by a protoplanetary disk detected in the mid- to far-infrared by the VISION survey, but not always detected and resolved at 225~GHz with ALMA \citep{van_Terwisga_2022}. Such a system has never been studied in the literature, and such an analysis would allow us to better understand the effects of the gravitational interaction between the companions on the structure and evolution of the disks, and to better constrain the typical tidal interaction that occurs in a multiple system.
\end{enumerate}

\begin{acknowledgements}
    The authors would like to thank the support of the Italian National Institute of Astrophysics (INAF) through the INAF GTO Grant \textit{ERIS \& SHARK GTO data exploitation} and the European Union's Horizon 2020 research and innovation program and the European Research Council via the ERC Synergy Grant \textit{ECOGAL} (project ID 855130). This paper makes use of the following ALMA data: ADS/JAO.ALMA$\#$2019.1.01813.S. ALMA is a partnership of ESO (representing its member states), NSF (USA) and NINS (Japan), together with NRC (Canada), NSTC (U.S.) ASIAA (Taiwan), and KASI (Republic of Korea), in cooperation with the Republic of Chile. The Joint ALMA Observatory is operated by ESO, AUI/NRAO and NAOJ. We also want to sincerely thank Steven N. Shore and Riccardo Franceschi for helpful discussions, Stefano Rinaldi for reading the first draft of this article. 
\end{acknowledgements}

\bibliographystyle{aa}
\bibliography{sodaiii}

\begin{appendix} 
\FloatBarrier
\section{Panels showing the multiple systems in ALMA and VISION (JHKs) data} \label{appendixC}
In this section, we present some continuum ALMA images at 225~GHz from the SODA survey \citep{van_Terwisga_2022} in which our multiple systems are visible. For each image, the source at the center indicates the primary disk, i.e. the brightest disk in the multiple system according to JHKs magnitudes in VISION III \citep{Grosschedl19}. 

\begin{figure} [hbt!]
\centering
   \includegraphics[width=\columnwidth]{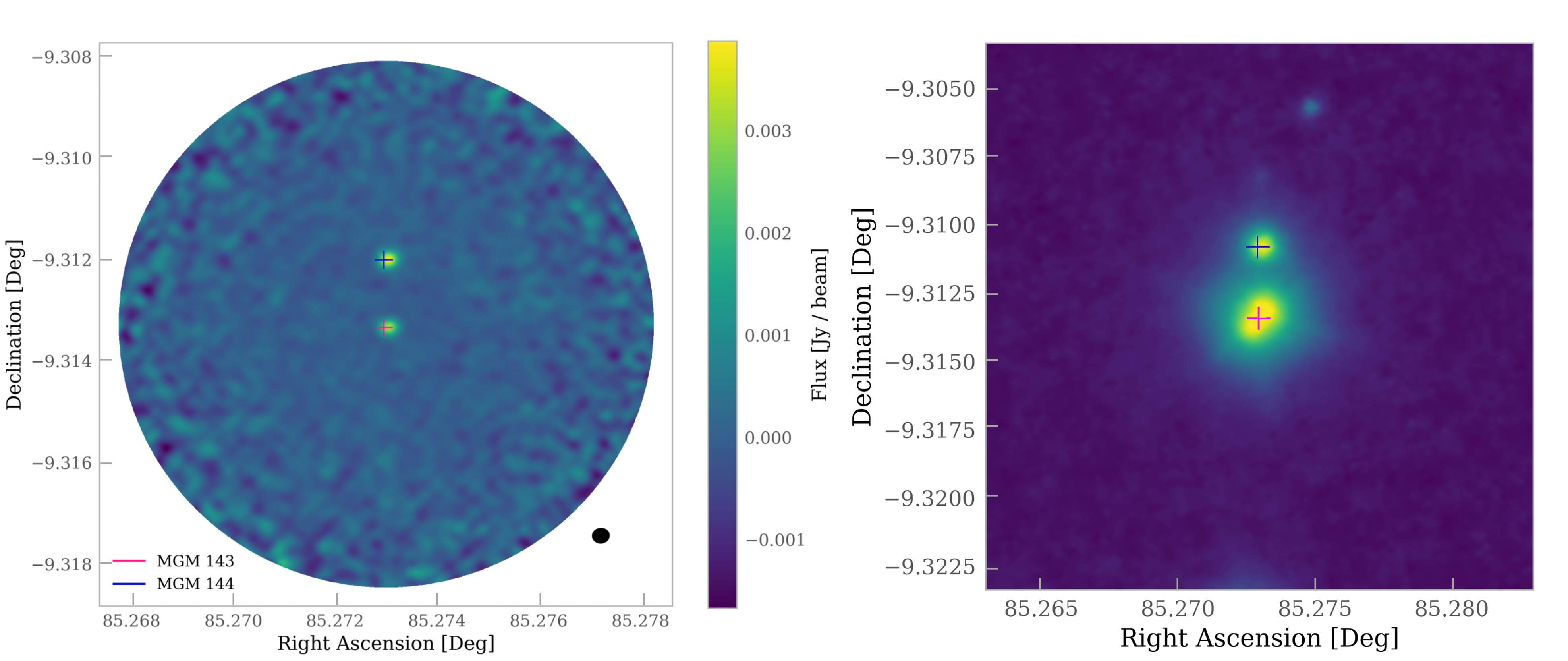}
     \caption{Continuum ALMA image at 225~GHz from the SODA survey \citep{van_Terwisga_2022}, binary system MGM~143 - MGM~144. The crosses mark the positions of the sources within the multiple system, identified by the index in either the VISION~III catalog \citep{Grossschedl20} or in the \cite{Megeath12} catalog. The beam is shown in black in the right-hand.}
     
\end{figure}

\begin{figure} [hbt!]
\centering
   \includegraphics[width=\columnwidth]{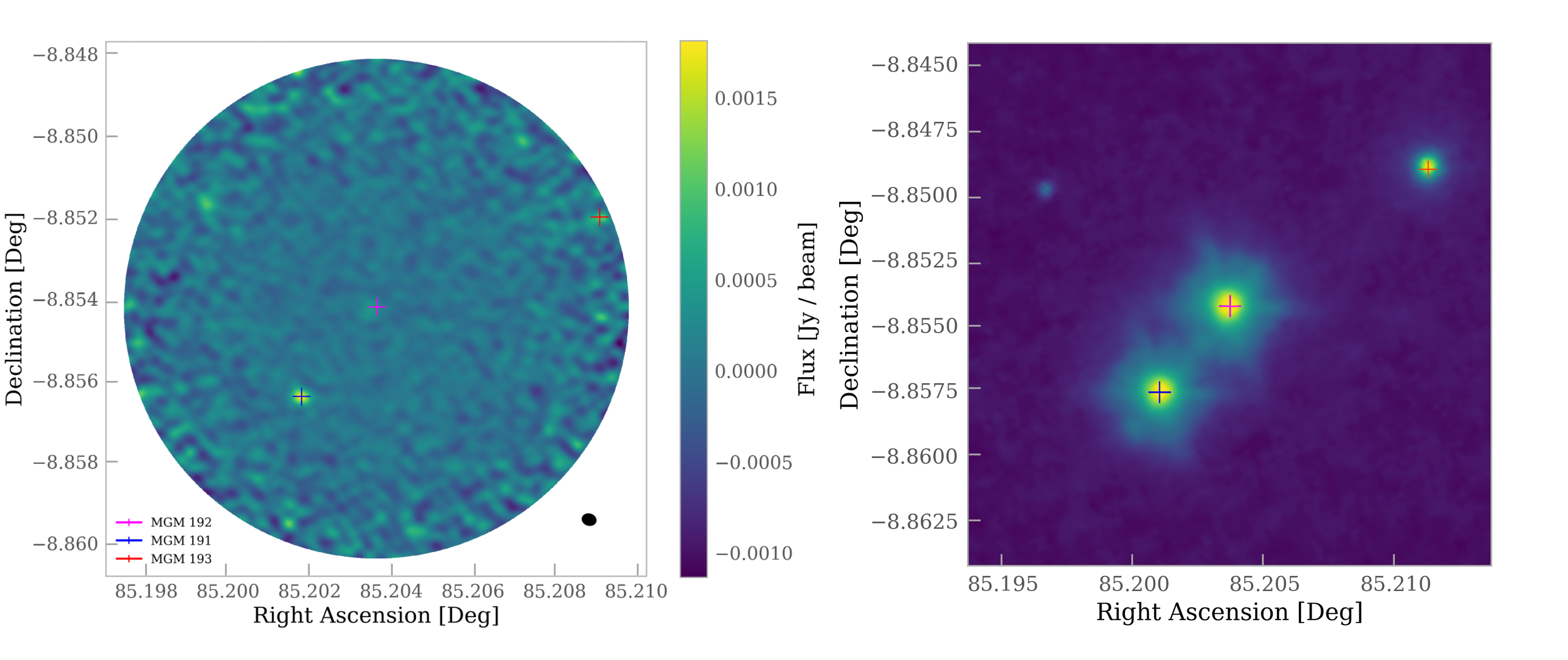}
     \caption{Panels and symbols are similar to Fig.~A.1. Triple system MGM~192 - MGM~191 - MGM~193.}
     
\end{figure}

\vspace{-100pt}
\begin{figure}
\centering
   \includegraphics[width=\columnwidth]{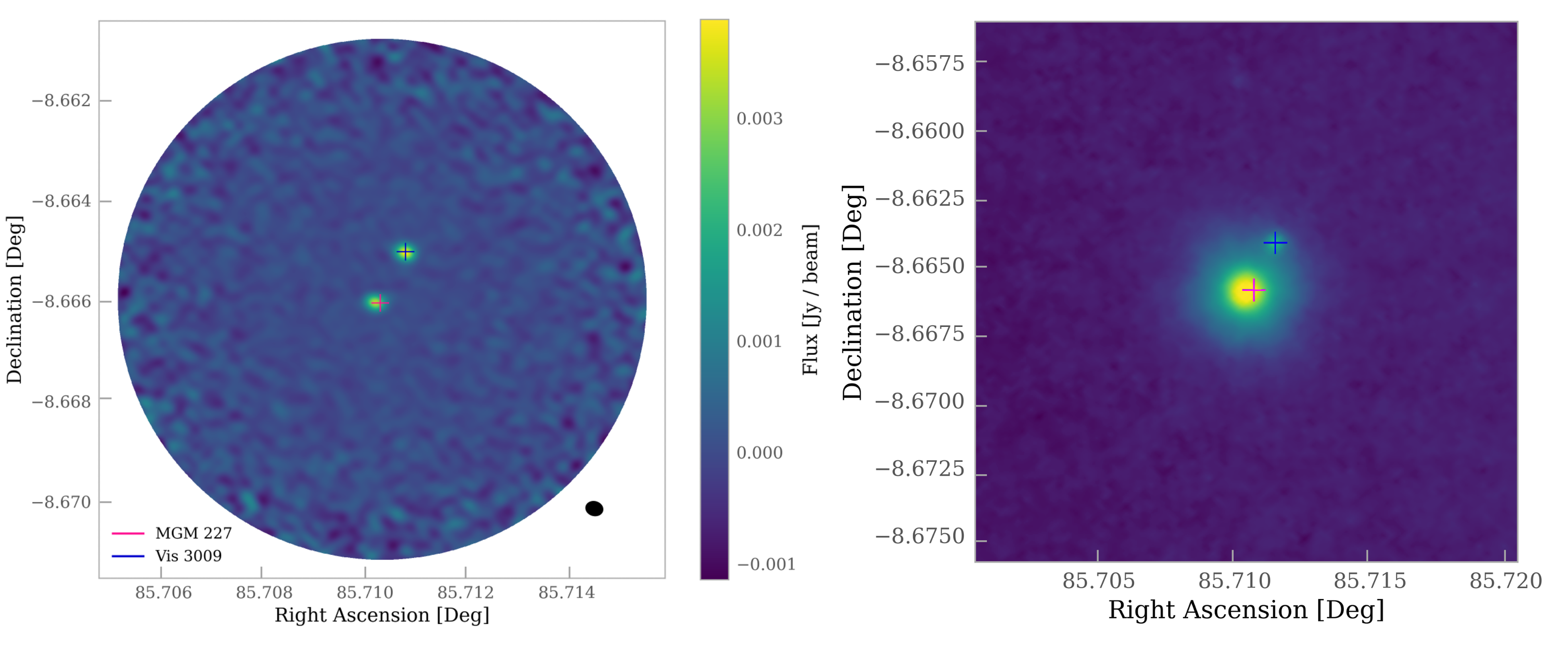}
     \caption{Panels and symbols are similar to Fig.~A.1. Binary system MGM~227 - Vis~3009.}
     
\end{figure}

\begin{figure}
\centering
   \includegraphics[width=\columnwidth]{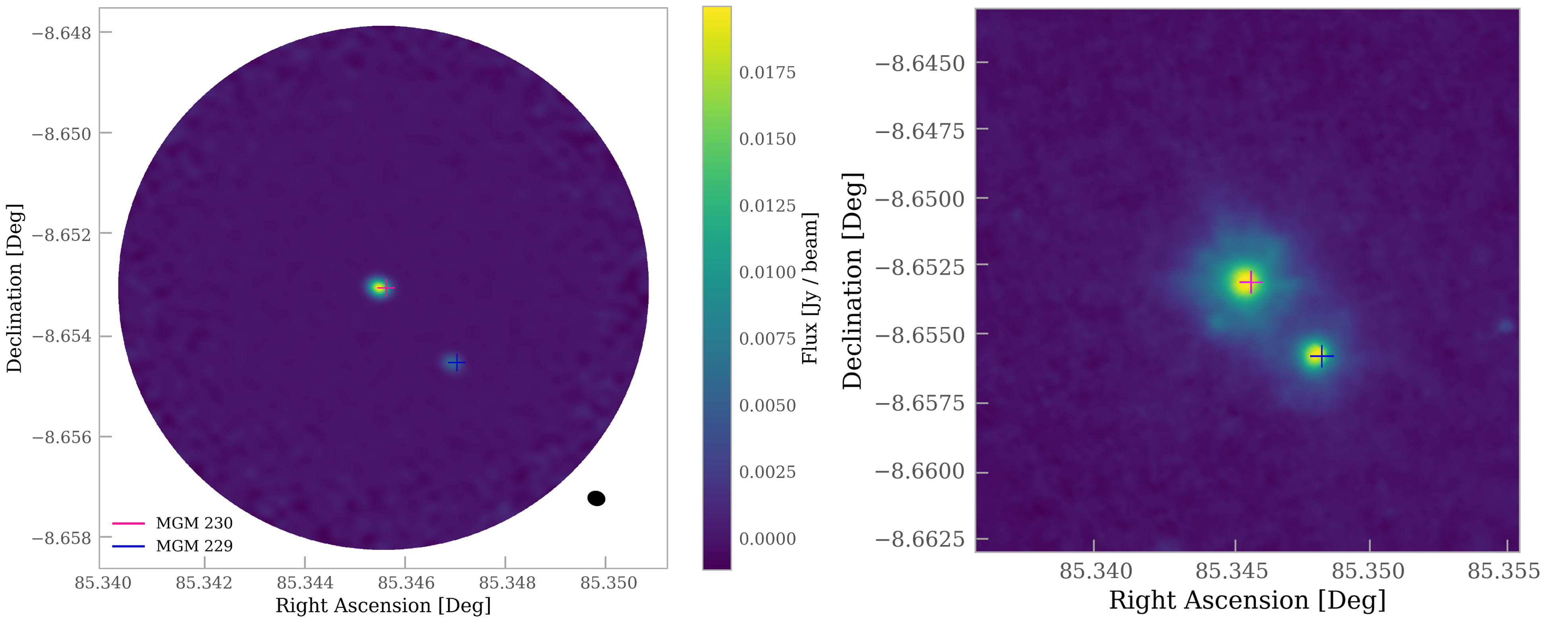}
     \caption{Panels and symbols are similar to Fig.~A.1. Binary system MGM~230 - MGM~229.}
     
\end{figure}

\begin{figure}
\centering
   \includegraphics[width=\columnwidth]{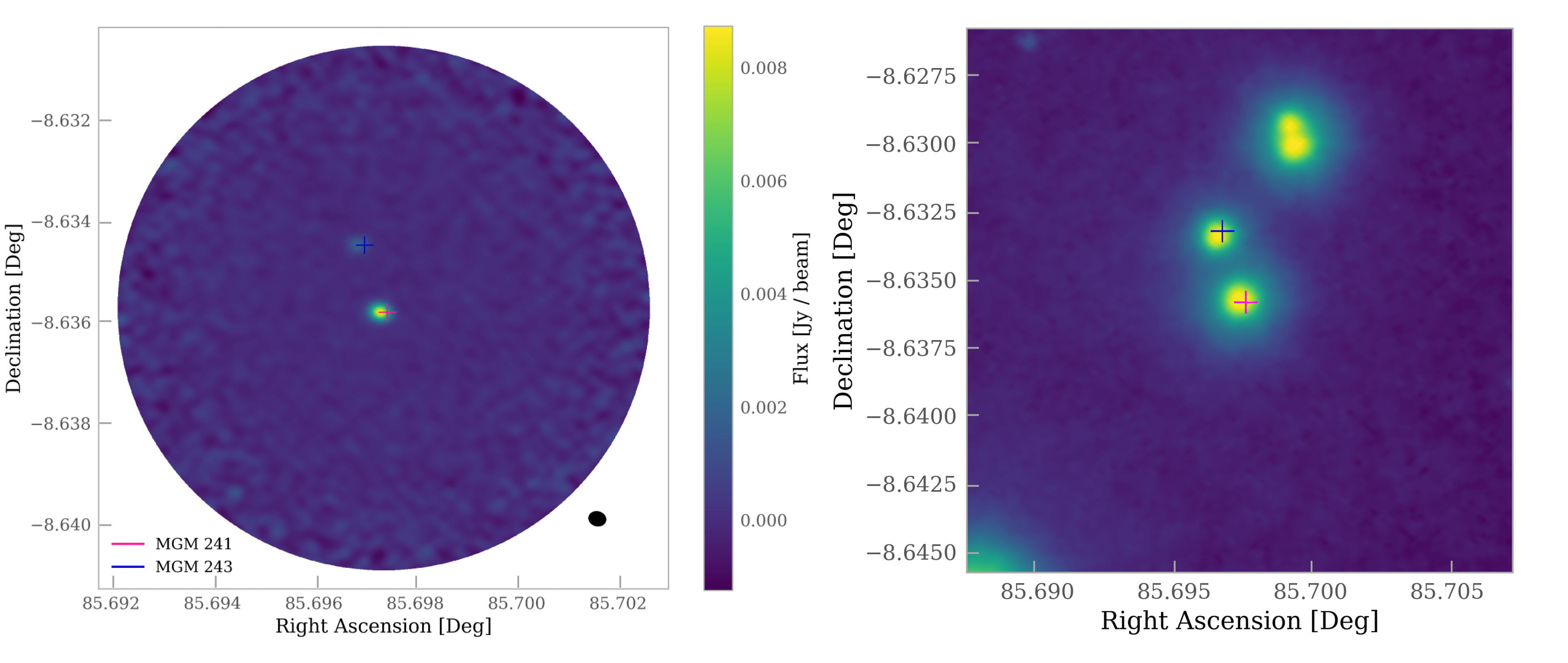}
     \caption{Panels and symbols are similar to Fig.~A.1. Binary system MGM~241 - MGM~243.}
     
\end{figure}

\begin{figure}
\centering
   \includegraphics[width=\columnwidth]{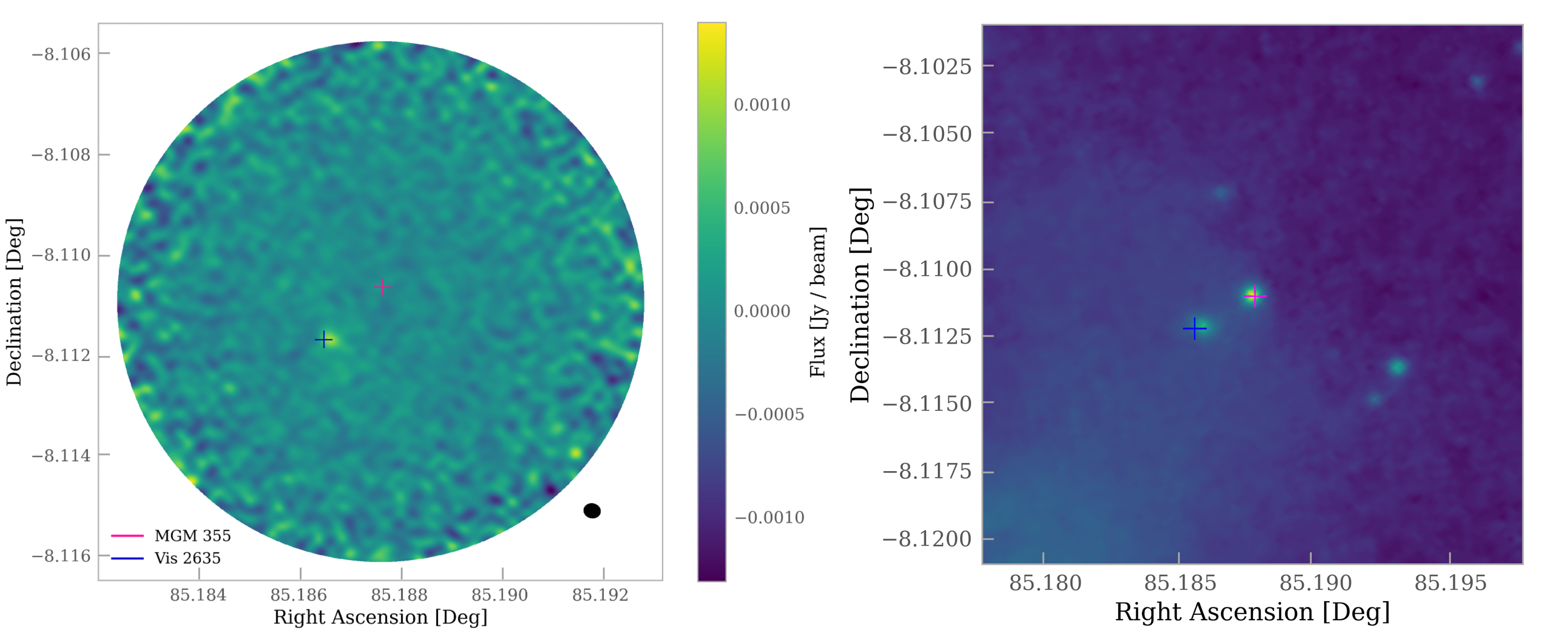}
     \caption{Panels and symbols are similar to Fig.~A.1. Binary system MGM~355 - Vis~2635.}
     
\end{figure}

\begin{figure}
\centering
   \includegraphics[width=\columnwidth]{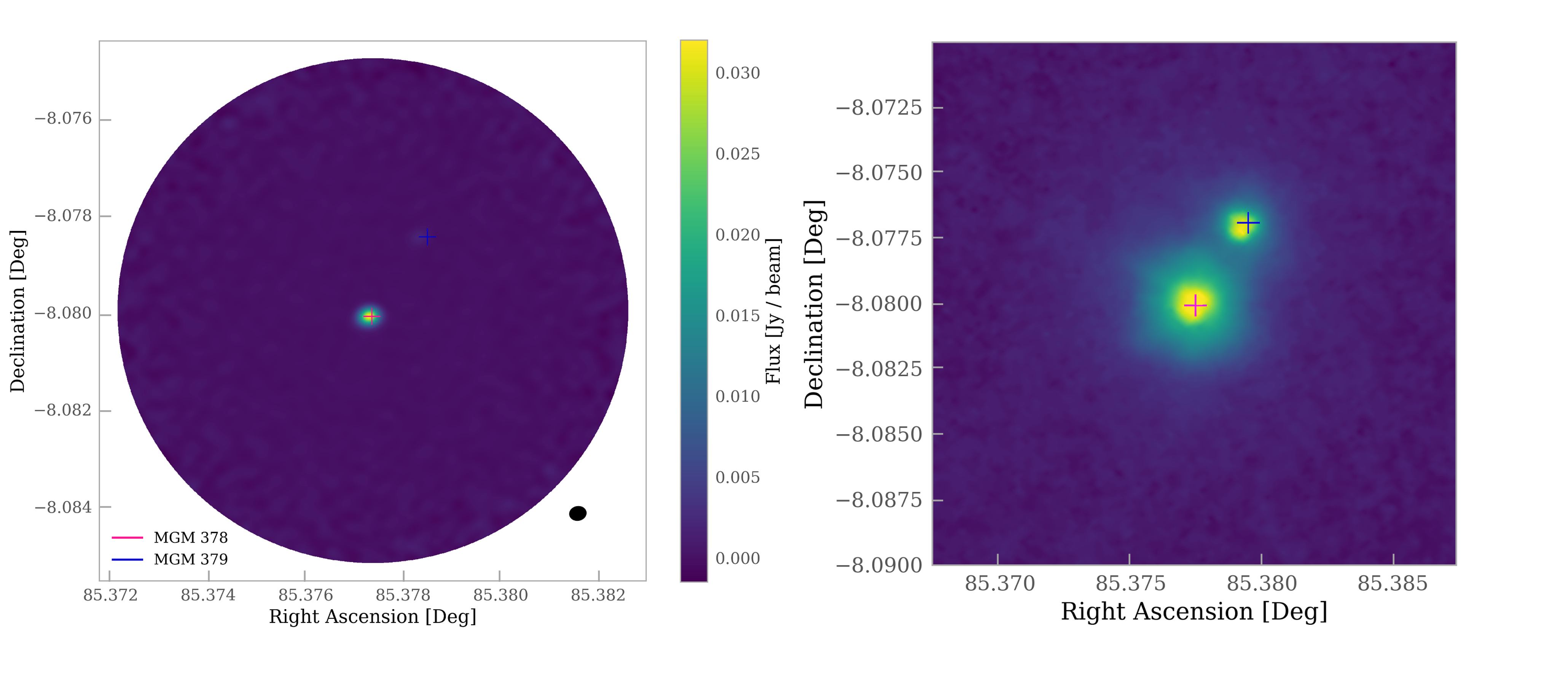}
     \caption{Panels and symbols are similar to Fig.~A.1. Binary system MGM~378 - MGM~379.}
     
\end{figure}

\begin{figure}
\centering
   \includegraphics[width=\columnwidth]{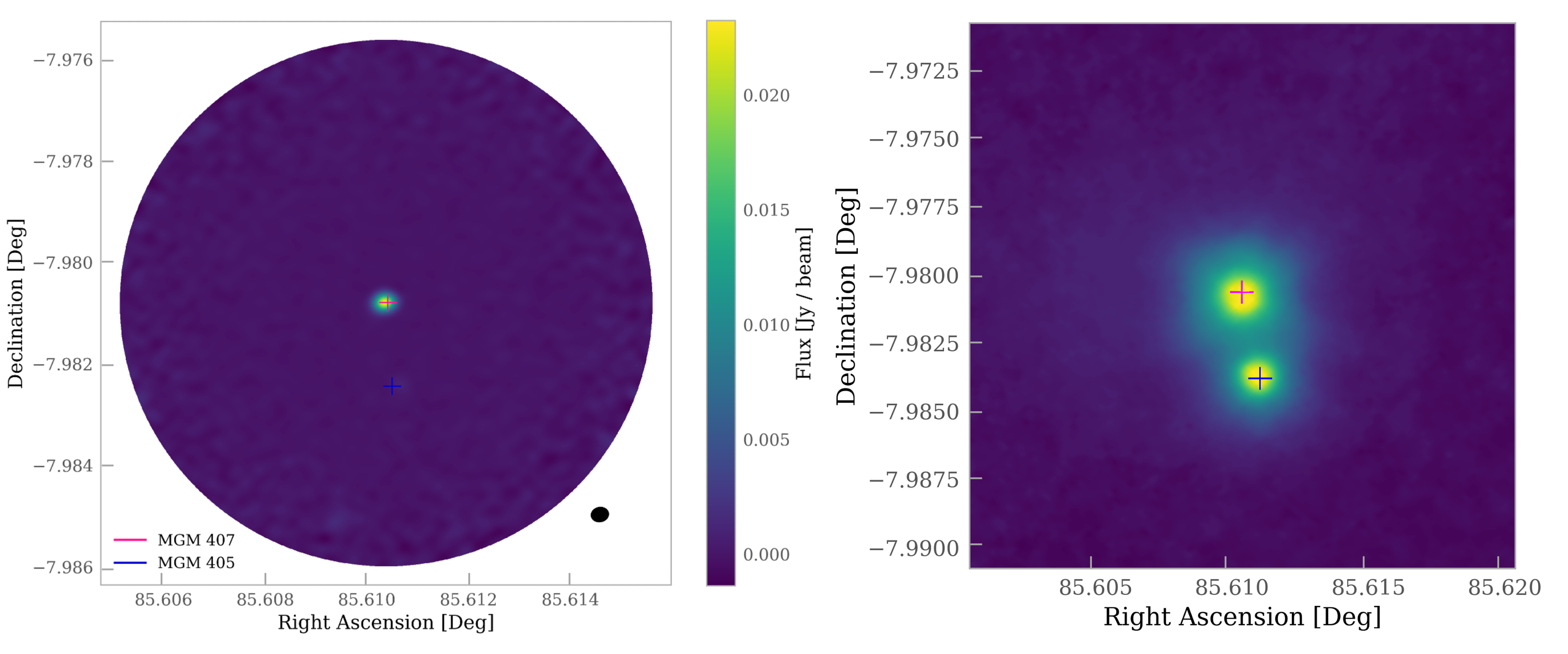}
     \caption{Panels and symbols are similar to Fig.~A.1. Binary system MGM~407 - MGM~405.}
     
\end{figure}

\begin{figure}
\centering
   \includegraphics[width=\columnwidth]{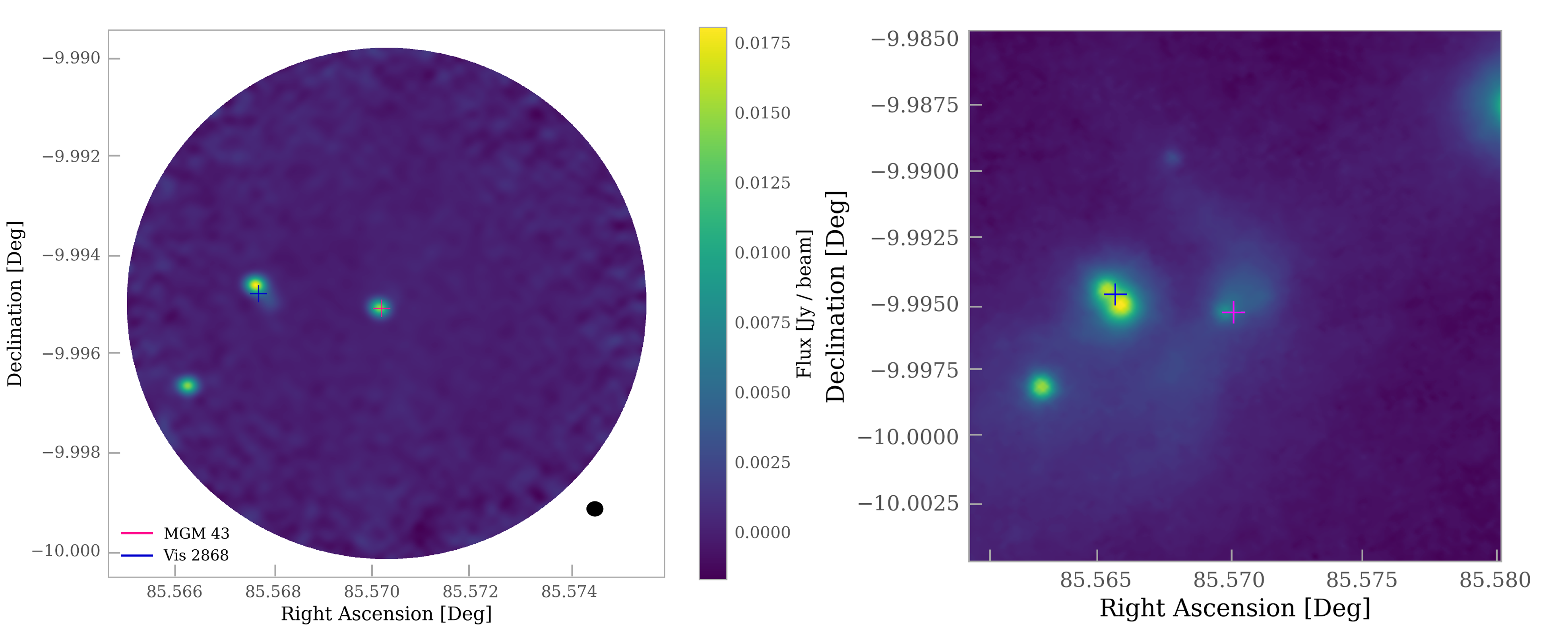}
    \caption{Panels and symbols are similar to Fig.~A.1. Binary system MGM~43 - Vis~2868.}
   \label{figure:MGM43}
\end{figure} 

\begin{figure}
\centering
   \includegraphics[width=\columnwidth]{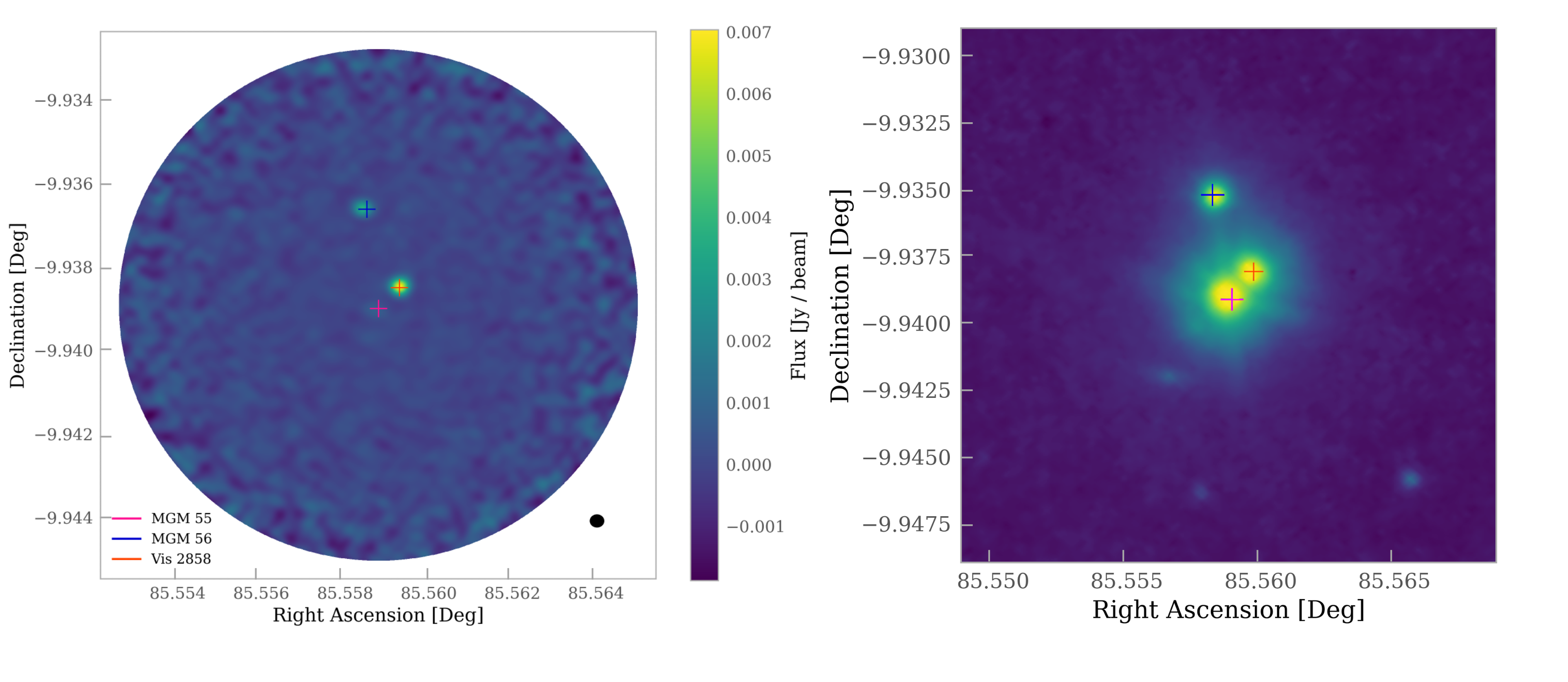}
     \caption{Panels and symbols are similar to Fig.~A.1. Triple system MGM~55 - MGM~56 - Vis~2858.}
     
\end{figure}

\begin{figure}
\centering
   \includegraphics[width=\columnwidth]{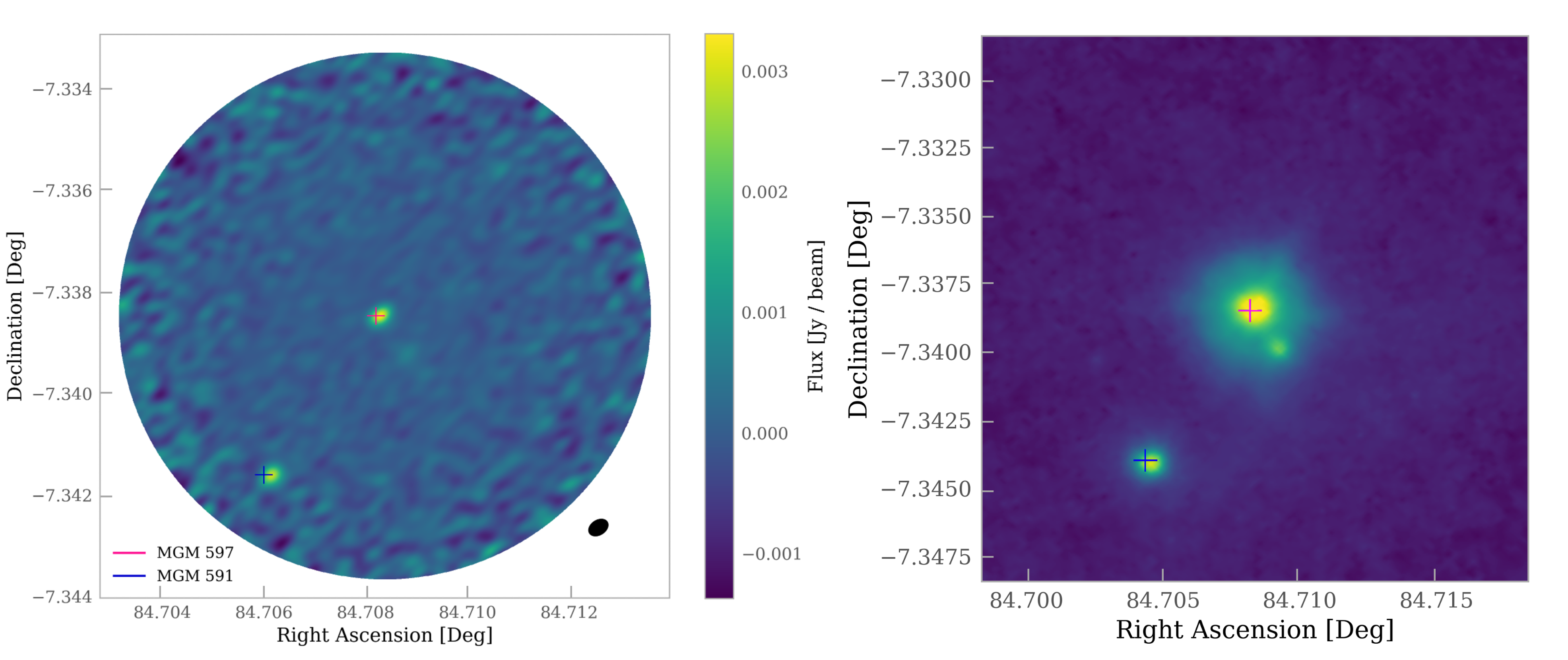}
     \caption{Panels and symbols are similar to Fig.~A.1. Binary system MGM~597 - MGM~591.}
     
\end{figure}

\begin{figure}
\centering
   \includegraphics[width=\columnwidth]{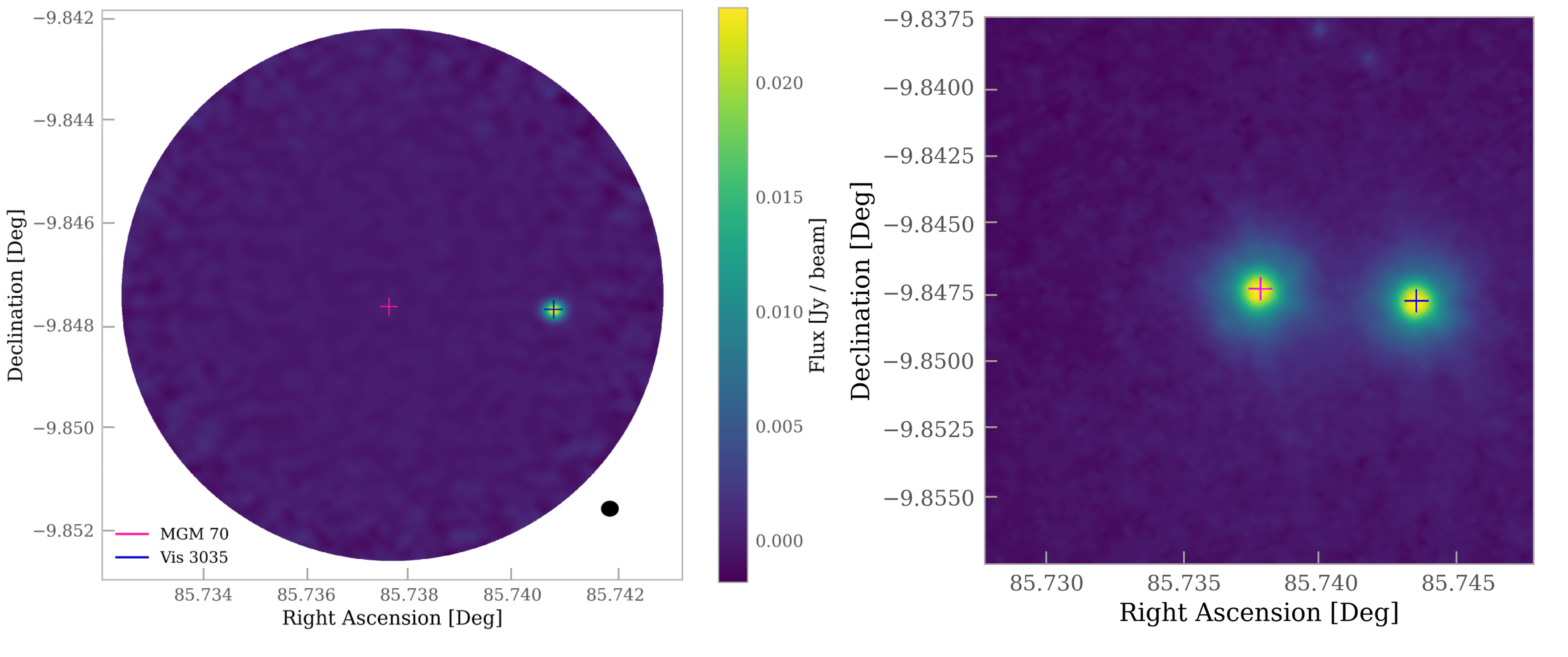}
     \caption{Panels and symbols are similar to Fig.~A.1. Binary system MGM~70 - Vis~3035.}
     
\end{figure}

\begin{figure}
\centering
   \includegraphics[width=\columnwidth]{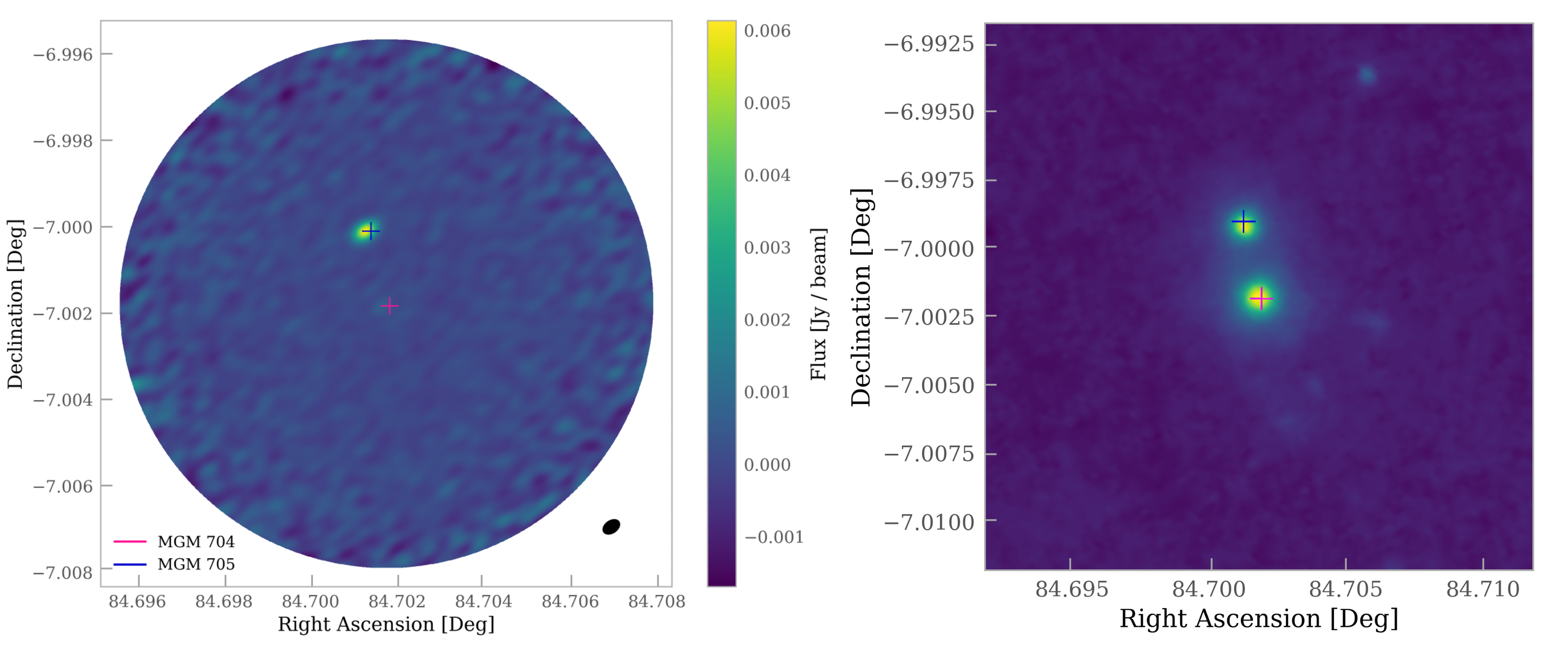}
     \caption{Panels and symbols are similar to Fig.~A.1. Binary system MGM~704 - MGM~705.}
     
\end{figure}

\begin{figure}
\centering
   \includegraphics[width=\columnwidth]{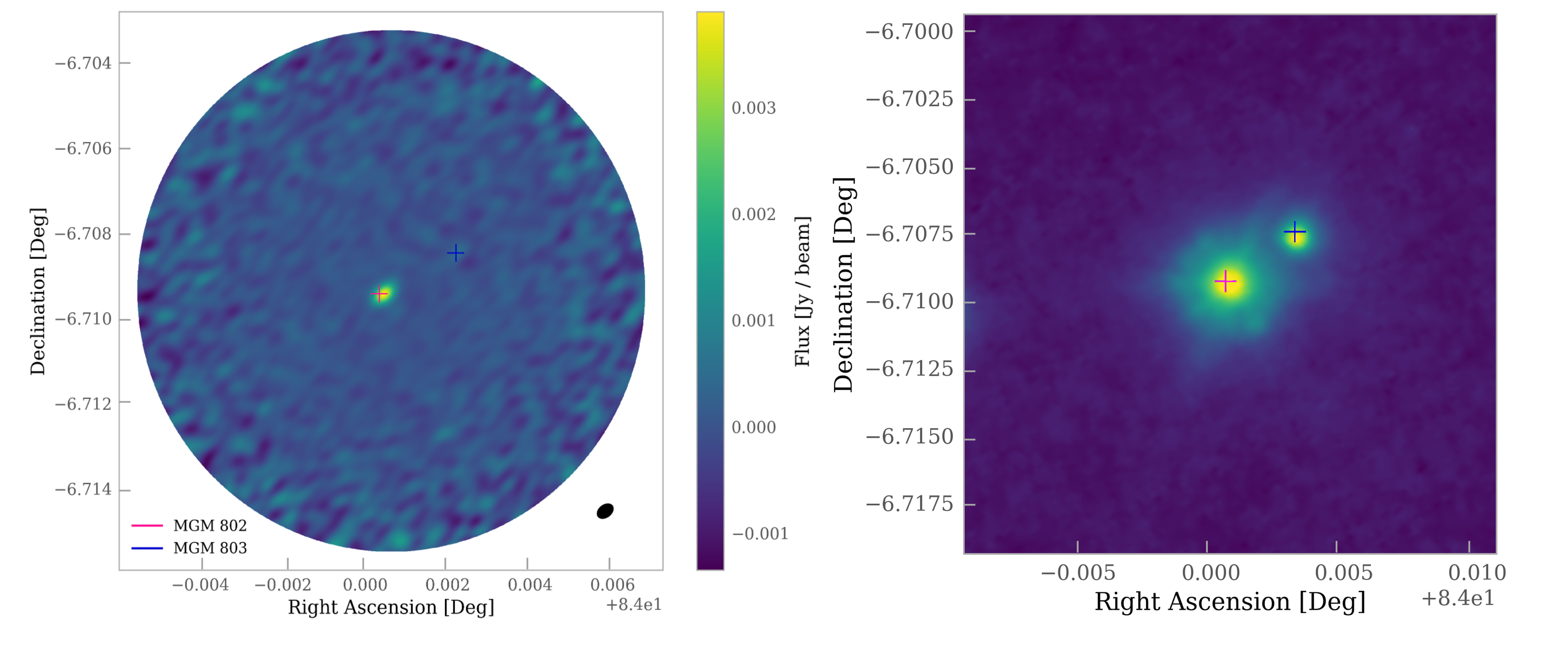}
     \caption{Panels and symbols are similar to Fig.~A.1. Binary system MGM~802 - MGM~803.}
     
\end{figure}

\begin{figure}
\centering
   \includegraphics[width=\columnwidth]{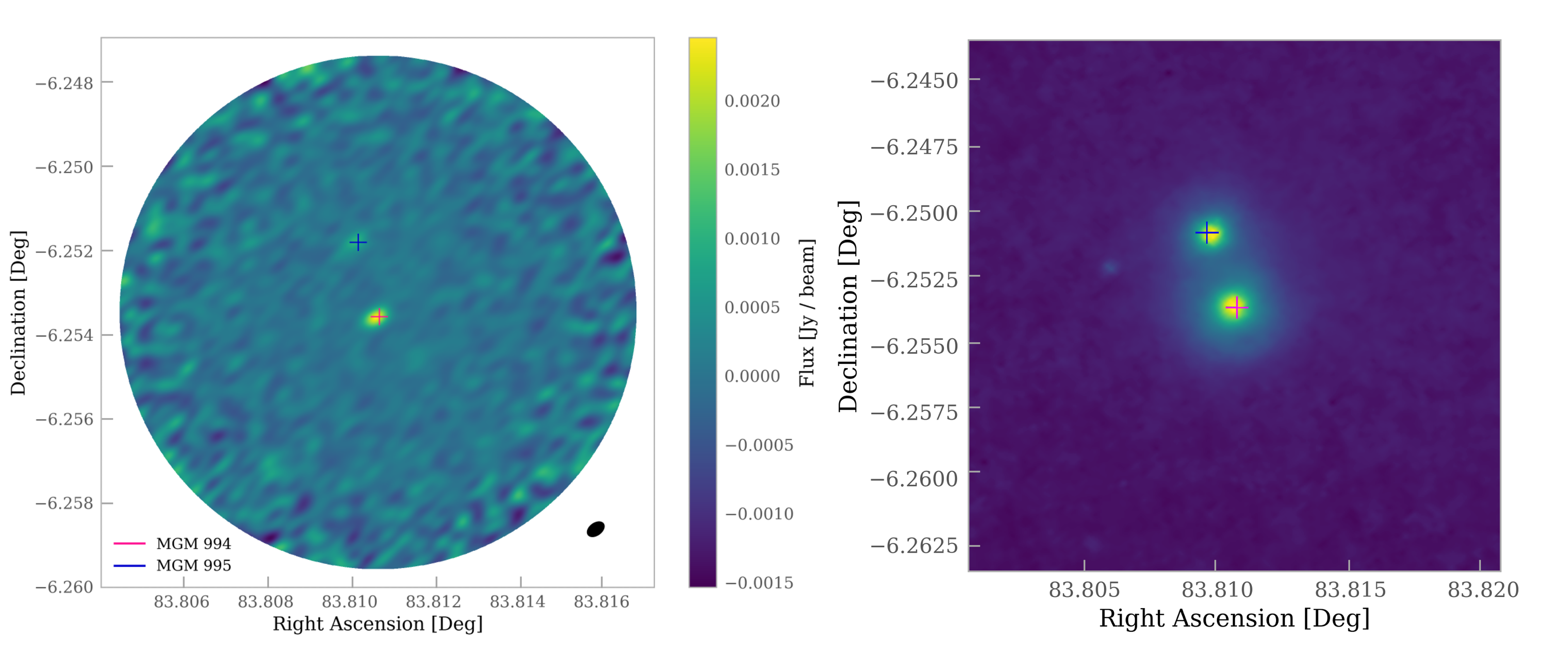}
     \caption{Panels and symbols are similar to Fig.~A.1. Binary system MGM~994 - MGM~995.}
     
\end{figure}

\begin{figure}
\centering
   \includegraphics[width=\columnwidth]{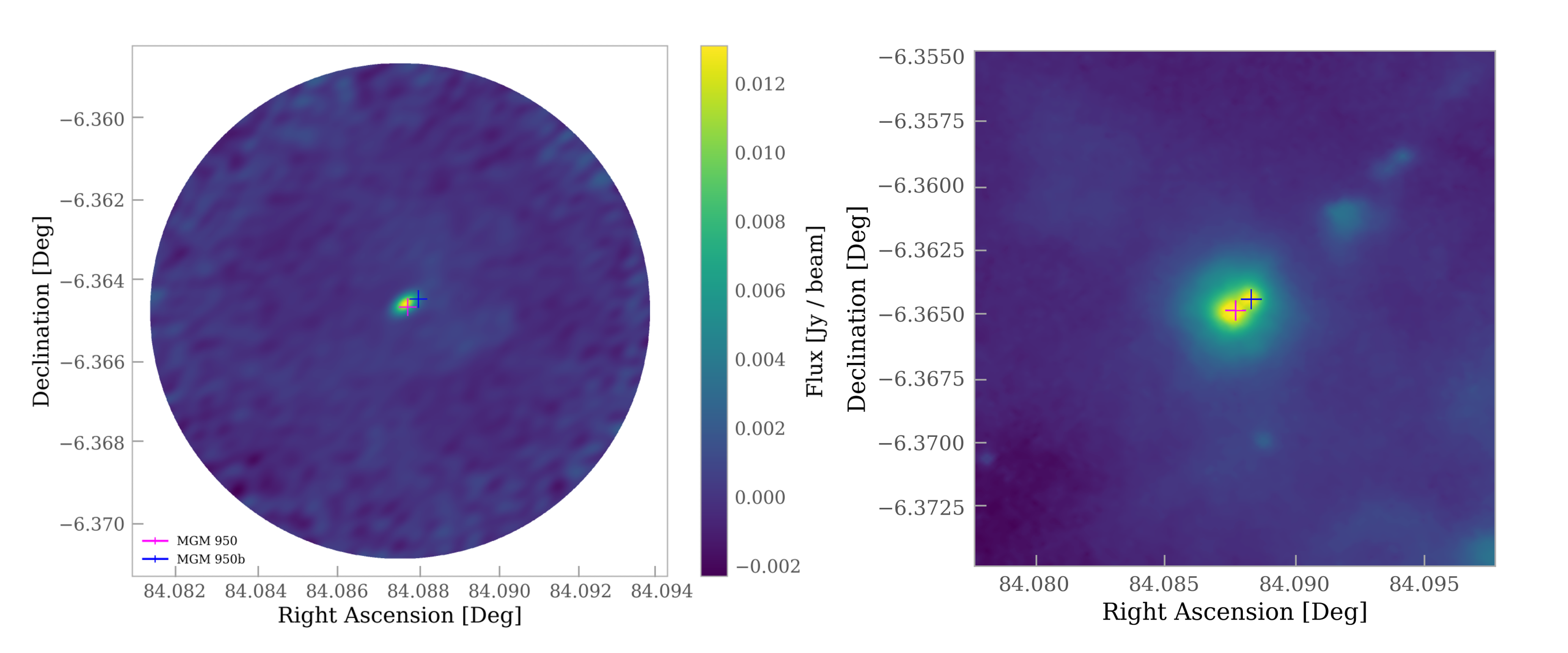}
    \caption{Panels and symbols are similar to Fig.~A.1. Binary system MGM~950 - MGM~950b.}
   \label{figure:MGM950}
\end{figure}

\begin{figure}
\centering
   \includegraphics[width=\columnwidth]{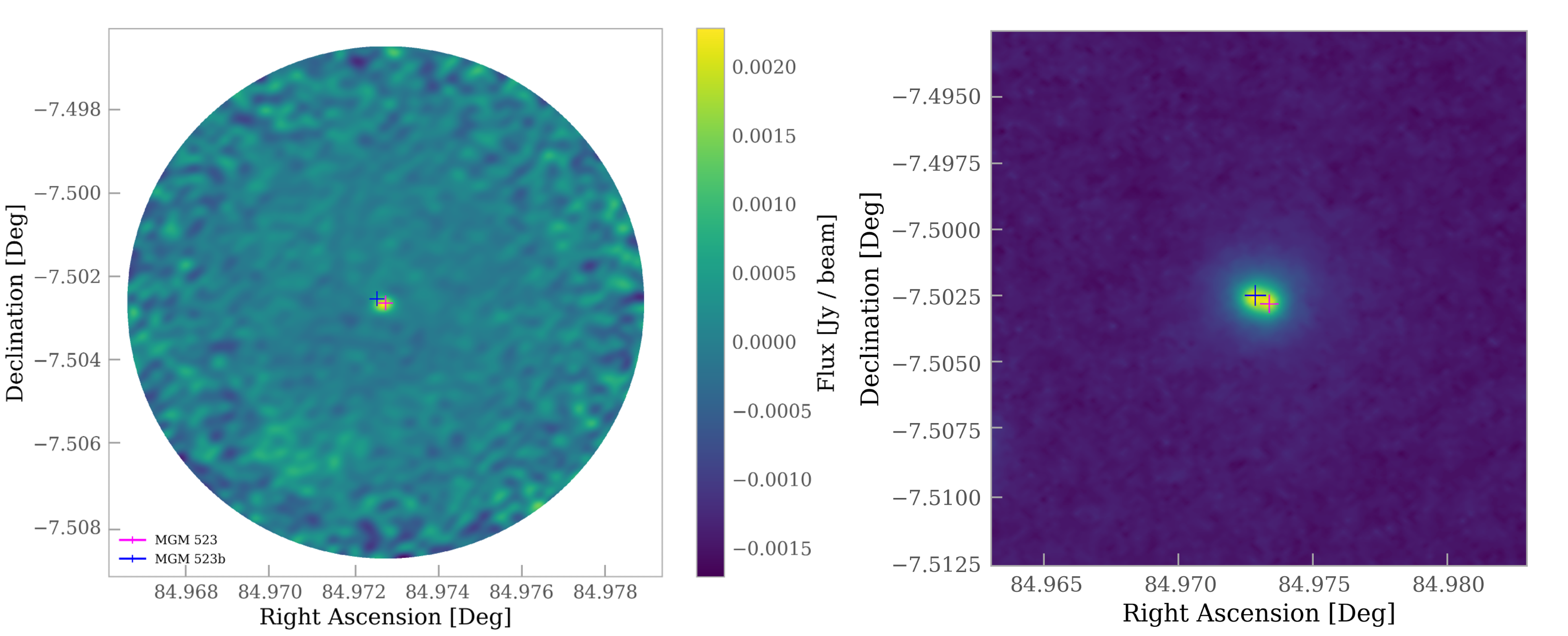}
    \caption{Panels and symbols are similar to Fig.~A.1. Binary system MGM~523 - MGM~523b.}
    \label{figure:MGM523}
\end{figure}

\FloatBarrier
\twocolumn
\section{System Separation} \label{appendixA}

This work focuses mainly on the dust mass of the Class II disk, which is directly proportional to the millimeter flux of the disk itself. Among the analyses performed, special attention has been given to deriving the relationship between the system mean mass, the system flux, and the system separation. \par
\begin{figure}[hbt!]
\centering
   \includegraphics[width=\columnwidth]{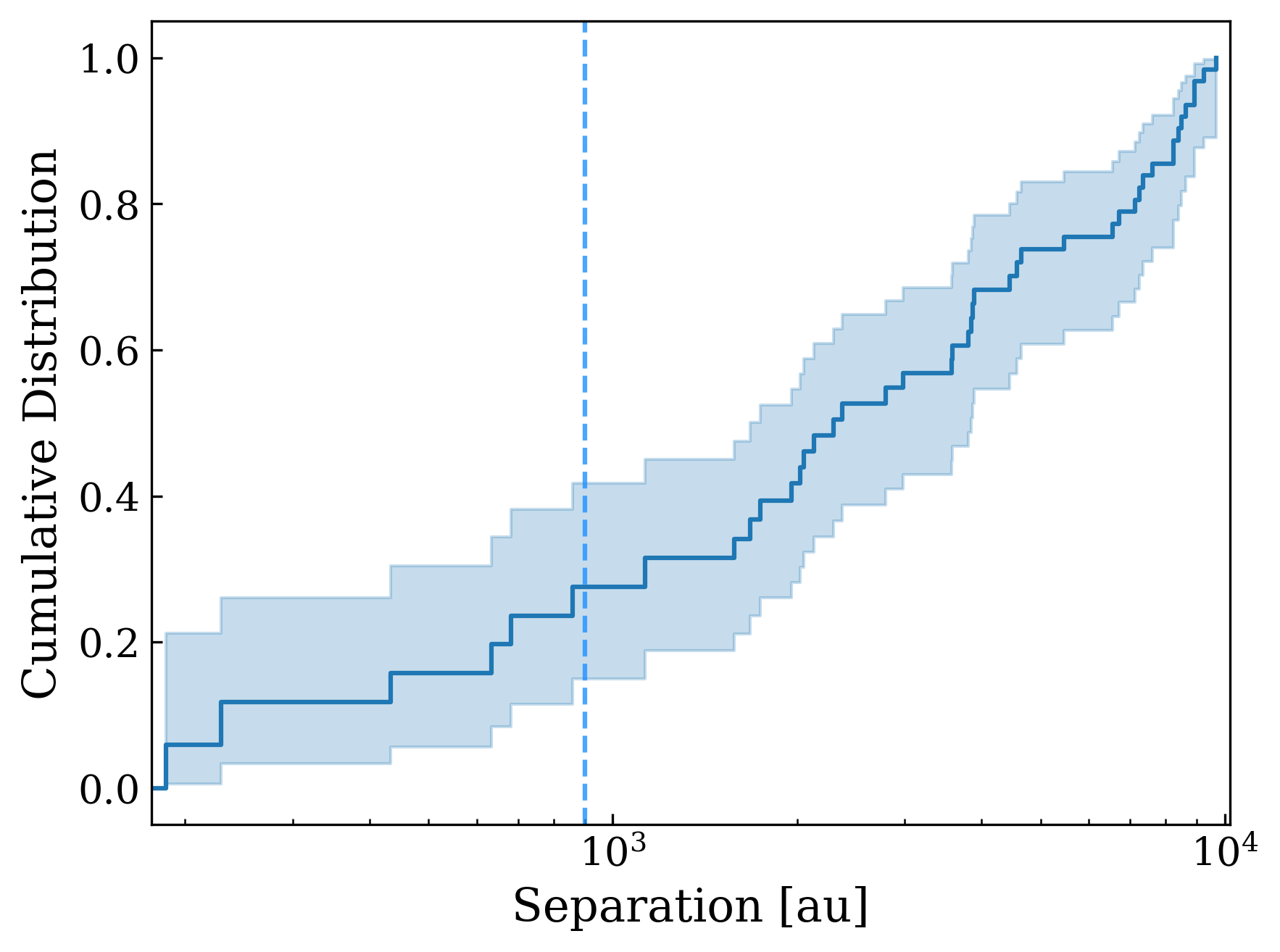}
     \caption{Cumulative distribution of the system separation inferred with the Kaplan-Meier estimator for binary systems in L1641 and L1647, Orion~A. The closer systems resolved by \cite{Kounkel16} are located before the blue vertical line.}
     \label{figure:sepcumulative}
\end{figure}

In this appendix, we focus on the separation of multiple systems. Figure~\ref{figure:sepcumulative} shows the cumulative distribution of the projected separation of our binary systems. Closer binary systems (projected separation between 100 and 1000~au) resolved by \cite{Kounkel16}, were also considered. 

\begin{figure}[hbt!]
\centering
   \includegraphics[width=\columnwidth]{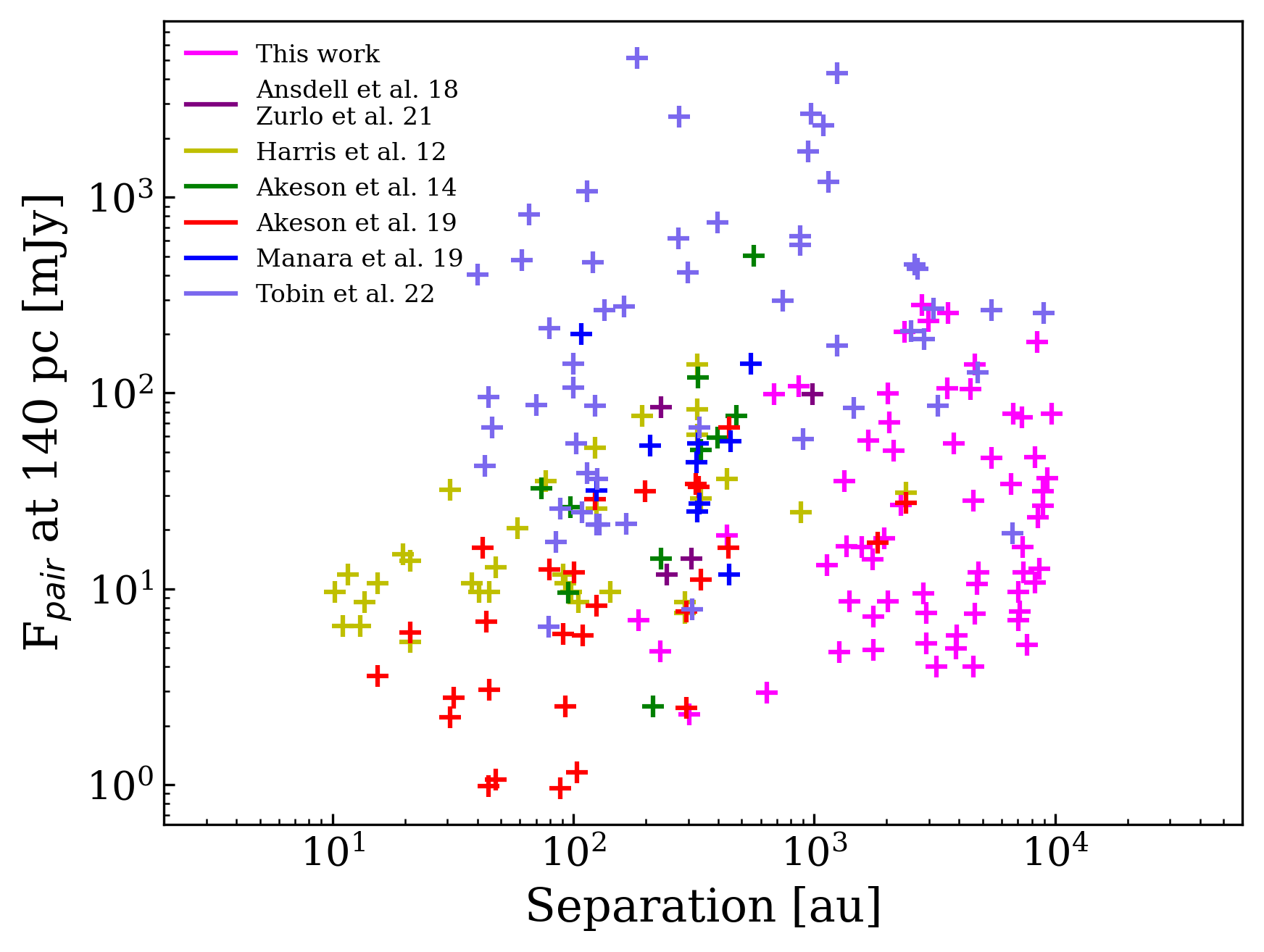}
     \caption{Millimeter flux in binary systems as a function of their projected separation, located in Orion, Lupus, and Taurus. The flux of each system is re-scaled at 140~pc, i.e. the average distance of Taurus. -~Class II~sources: Orion~A (this work), Lupus \citep{Zurlo21, Ansdell18}, and Taurus \citep{Akeson19, Manara19, Akeson14, Harris_2012}. -~Class 0, I, and FT~sources: Orion \citep{Tobin_2022}. We only consider systems detected by ALMA (0.87~mm).}
     \label{figure:fluxsep1}
\end{figure}

\begin{figure}[hbt!]
\centering
   \includegraphics[width=\columnwidth]{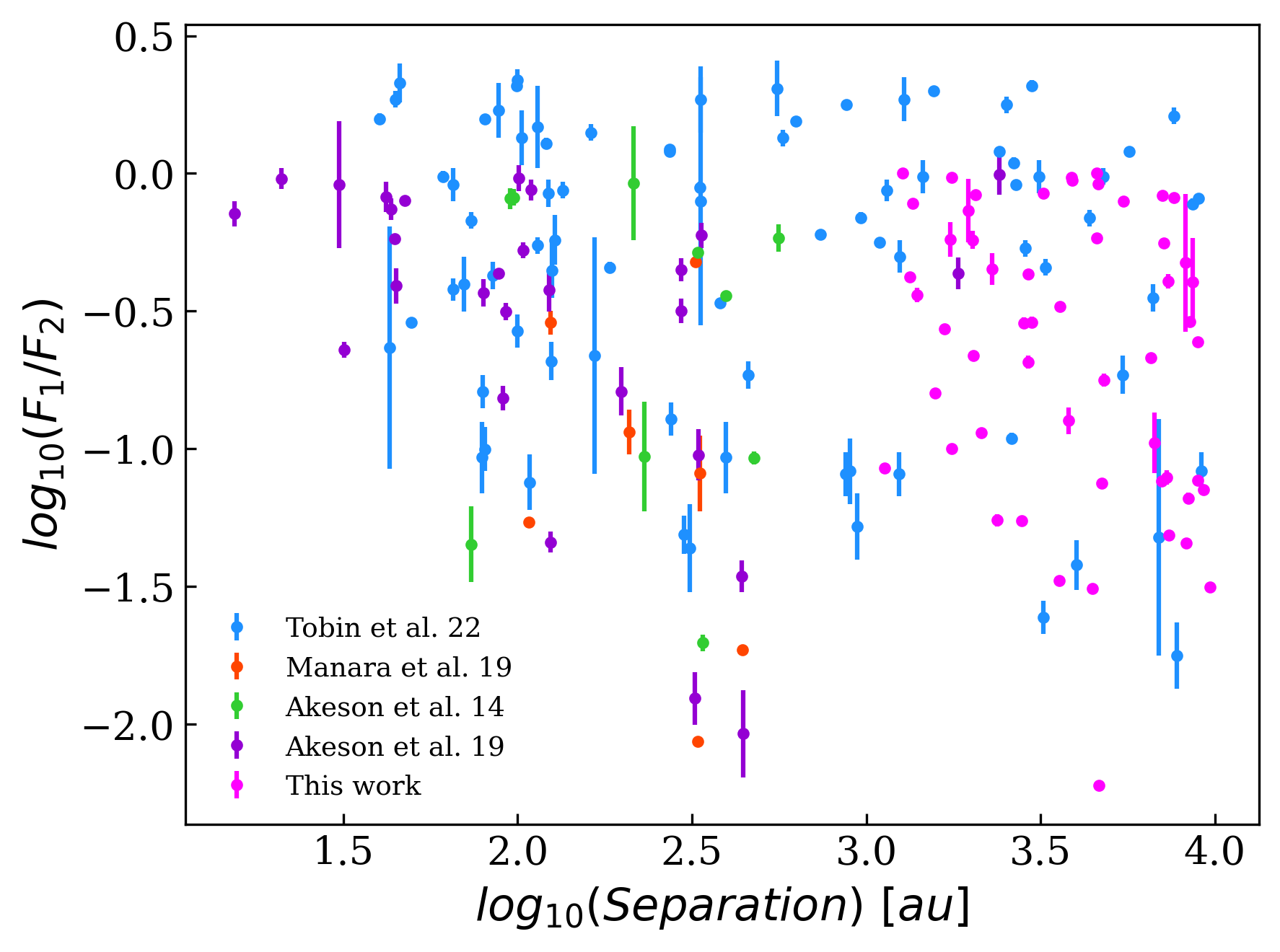}
     \caption{Flux density ratio vs. separation for binaries in Orion and Taurus. - Orion: Sample of multiple Class 0, I and FS disks by \cite{Tobin_2022}, and our multiple Class II disks. - Taurus: Multiple systems presented by \cite{Akeson19}, \cite{Manara19}, and \cite{Akeson14}.}
     \label{figure:tobinzagaria}
\end{figure}

Since the separation between companions can provide information about the formation process of the multiple system, it is interesting to consider objects of different ages. Our catalog only includes Class II disks, as well as the observations we have considered in our analyses for Taurus and Lupus refer only to Class II YSOs. \cite{Tobin_2022} characterize protostellar multiplicity in Orion considering younger and more embedded sources: Class 0, I, and Flat Spectrum (FT) YSOs. Considering all the binary systems in these surveys, in Figure~\ref{figure:fluxsep1} we again show the pair millimeter flux as a function of the system separation, while in Figure~\ref{figure:tobinzagaria} we show the flux density ratio over systems separation for the binary disk systems located in Orion and Taurus. Please note that in Figure~\ref{figure:fluxsep1}, the flux of each system is corrected for the frequency difference and re-scaled at the same distance. Indeed, the Class II sample is observed at 1.3~mm, while \cite{Tobin_2022} have observations at 0.85 or 9~mm. To convert the fluxes into 1.3~mm, we assume F$_{\nu} \propto \nu^{-3}$, valid under the assumption of optically thin emission with a dust opacity spectral index $\beta$ $\simeq$ 1. To re-scale the fluxes at 140~pc, we assume F$~\propto d^{-2}$. In Figure~\ref{figure:tobinzagaria}, instead, the flux ratio considered is simply equal to the ratio between the minor and major fluxes. This is because \cite{Tobin_2022} does not denote the primary and secondary disks within their binary systems.

Figure~\ref{figure:fluxsep1} shows that the pair flux of embedded double systems tends to be higher. Although the envelope of gas and dust surrounding these YSOs is more prominent at submillimeter and beyond wavelengths, it can still contribute to the millimeter-wavelength emission and hence to the observed millimeter flux. However, according to Figure~\ref{figure:tobinzagaria}, if we look at the flux density ratio, there is no obvious distinction between the different stages. The plot is uniformly populated and there is no correlation between the flux ratio and the separation.

\newpage
\FloatBarrier
\section{Flux-separation correlation: Corner plots} \label{AppendixCorn}
In this appendix we show the corner plots referring to Figures~\ref{figure:fluxsep} and \ref{figure:linearconst}

\begin{figure}[hbt!]
\centering
   \includegraphics[width=\columnwidth]{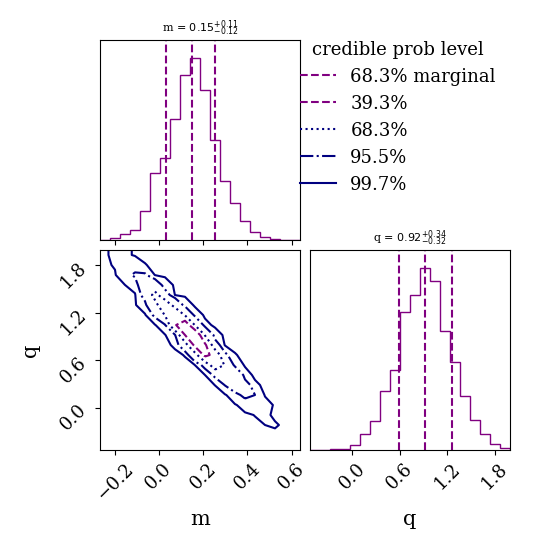}
    \caption{Referring to Figure~\ref{figure:fluxsep}, corner plot for the posterior parameter distribution, and the correlations between the free parameters m (slope) and q (intercept).}
     
\end{figure}

\begin{figure}[hbt!]
\centering
   \includegraphics[width=\columnwidth]{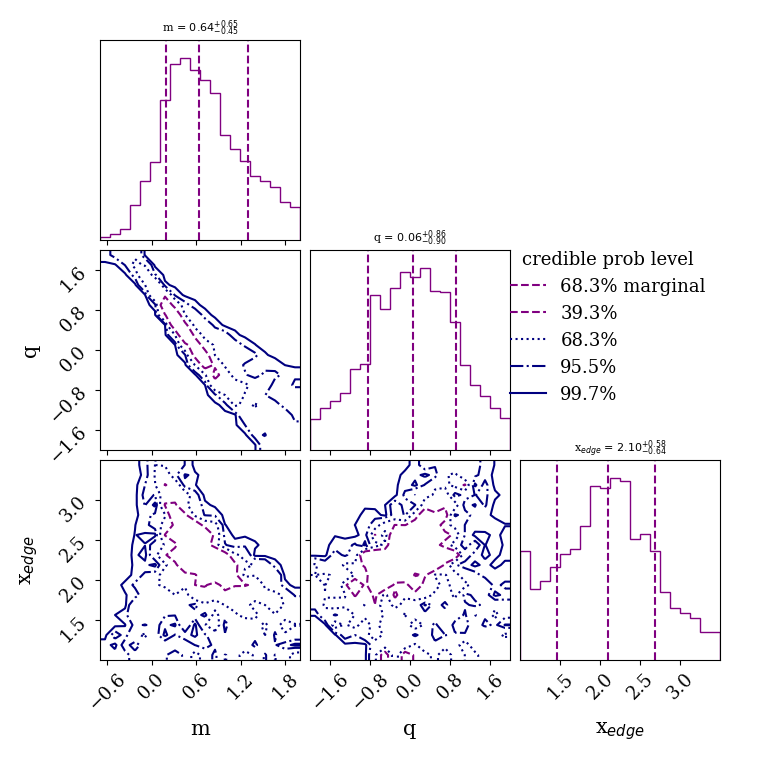}
     \caption{Referring to Figure~\ref{figure:linearconst}, corner plot for the posterior parameter distribution, and the correlations between the free parameters m (slope), q (intercept), and x$_{edge}$ (separation threshold).}
\end{figure}

\newpage
\FloatBarrier
\section{Additional tables} 

\onecolumn
\begin{longtable}{cccc}
\caption{Multiple systems in the L1641 and L1647 regions of Orion~A. In the 1st column, the 1st index indicates the Primary disk.}\\ 
\hline\hline 
 Multiple System & Separation & Mean Mass & Average  \\
 (Vision Index) & (au) & (M$_{\oplus}$) & Distance (pc)\\
\hline\hline 
\endfirsthead
\caption{continued}\\
\hline\hline
 Multiple System & Separation & Mean Mass & Average  \\
 (Vision Index) & (au) & (M$_{\oplus}$) & Distance (pc)\\
\hline\hline
\endhead
\hline
\endfoot
\hline
\endlastfoot
 1000 - 1006 &  2294.71 & 8.012 $\pm$ 0.25 & 384 \\
 1739 - 1745 &  1756.08 & 1.29 $\pm$ 0.36 & 384 \\
 1956 - 1954 &  8926.72 & 9.47 $\pm$ 0.25 & 384 \\
 2413 - 2414 &  2133.75 & 15.29 $\pm$ 0.25 & 386 \\
 2495 - 2494 &  2917.50 & 2.03 $\pm$ 0.36 & 398 \\
 2655 - 2657 &  1578.05 & 4.80 $\pm$ 0.35 & 404 \\
 2697 - 2698 &  2053.65 & 21.17 $\pm$ 0.91 & 441 \\
 2784 - 2787 &  2795.71 & 84.58 $\pm$ 0.28 & 405 \\
 2746 - 2743 &  2978.96 & 70.11 $\pm$ 1.50 & 422 \\
 2714 - 2715 &  2832.32 & 2.61 $\pm$ 0.32 & 433 \\
 2889 - 2888 &  2373.41 & 61.39 $\pm$ 0.28 & 405 \\
 2859 - 2683 - 2858 & 2647.81 & 7.48 $\pm$ 0.99 & 454 \\
 2982 - 2984 &  2024.88 & 29.864 $\pm$ 0.62 & 429 \\
 2955 - 2956 &  1269.40 & 1.09 $\pm$ 0.35 & 432 \\
 2960 - 2959 &  1958.48 & 5.45 $\pm$ 0.33 & 442 \\
 3054 - 3055 &  1744.14 & 4.23 $\pm$ 0.59 & 454 \\
 1130 - 1107 &  3810.64 & 16.52 $\pm$ 0.40 & 385 \\
 1508 - 1514 &  4453.69 & 31.38 $\pm$ 0.39 & 384 \\
 2142 - 2141 &  3578.75 & 31.56 $\pm$ 0.38 & 385 \\
 2303 - 2306 &  3875.99 & 1.47 $\pm$ 0.35 & 386 \\
 2497 - 2496 &  3894.07 & 1.73 $\pm$ 0.37 & 396 \\
 2650 - 2648 - 2642 & 3476.82 & 4.7 $\pm$ 0.33 & 428 \\
 2882 - 2880 &  3222.03 & 0.89 $\pm$ 0.30 & 405 \\
 2979 - 2987 &  4633.78 & 1.22 $\pm$ 0.31 & 405 \\
 3022 - 3024 &  7630.08 & 1.38 $\pm$ 0.56 & 454 \\
 1562 - 1548 &  7260.48 & 22.45 $\pm$ 0.38 & 384 \\
 2230 - 2229 &  7356.17 & 4.89 $\pm$ 0.40 & 386 \\
 2296 - 2295 &  5456.61 & 14.03 $\pm$ 0.38 & 386 \\
 2418 - 2420 &  6716.99 & 23.41 $\pm$ 0.36 & 393 \\
 2501 - 2502 &  7135.36 & 2.30 $\pm$ 0.38 & 396 \\
 2570 - 2570b - 2567 - 2572 & 4385.40 & 84.56 $\pm$ 0.81 & 404 \\
 2811 - 2812 &  7054.21 & 1.29 $\pm$ 0.30 & 404 \\
 3066 - 3069 &  6553.77 & 10.26 $\pm$ 0.52 & 454 \\
 3101 - 3100 &  8238.90 & 3.22 $\pm$ 0.37 & 430 \\
 2205 - 2209 &  8402.28 & 54.74 $\pm$ 0.36 & 384 \\
 2563 - 2565 &  8914.26 & 7.96 $\pm$ 0.34 & 403 \\
 2672 - 2680 &  8489.11 & 6.96 $\pm$ 0.28 & 403 \\
 2686 - 2680 &  8256.15 & 14.15  $\pm$ 0.29 & 403 \\
 2644 - 2642 &  9691.03 & 23.57 $\pm$ 0.32 & 428 \\
 2696 - 2701 &  8630.73 & 3.27 $\pm$ 0.31 & 403 \\
 2716 - 2707 &  9252.85 & 10.96 $\pm$ 0.55 & 434 \\
 2861 - 2862 &  1398.61 & 2.54 $\pm$ 0.30 & 405 \\
 2010 - 2009 &  2023.63 & 2.57 $\pm$ 0.36 & 384 \\
 2058 - 2051 &  1360.99 & 4.90 $\pm$ 0.26 & 384 \\
 2314 - 2313 &  1337.12 & 10.63 $\pm$ 0.38 & 385 \\
 2545 - 2544 &  1754.64 & 1.76 $\pm$ 0.34 & 399 \\
 2633 - 2635 &  1129.81 & 3.87 $\pm$ 0.42 & 404 \\
 2836 - 3050 &  1676.91 & 16.83 $\pm$ 0.43 & 429 \\
 3011 - 3009 &  2915.12 & 1.53 $\pm$ 0.51 & 430 \\
 3014 - 3013 &  4796.05 & 3.56 $\pm$ 0.35 & 429 \\
 2301 - 2304 &  3851.21 & 101.84 $\pm$ 0.35 & 386 \\
 2567 - 2572 &  4585.81 & 8.38  $\pm$ 0.57 & 403 \\
 2867 - 2868 &  3592.37 & 75.66 $\pm$ 0.33 & 405 \\
 2937 - 2940 &  4739.58 & 2.95 $\pm$ 0.76 & 421 \\
 3041 - 3035 &  4649.14 & 40.42 $\pm$ 0.41 & 454 \\
 2399 - 2391 &  7028.63 & 2.53 $\pm$ 0.66 & 434 \\
 2489 - 2487 &  7378.79 & 3.31 $\pm$ 0.37 & 454 \\
 2425 - 2429 - 2501 & 3856.21 & 7.94 $\pm$ 0.28 &  394  \\
 2513 - 2513b  & 433.0 & 11.18 $\pm$ 0.40 & 401 \\
 2518 - 2518b  & 634.0 & 1.76 $\pm$ 0.33 & 401 \\
 2492 - 2492b  & 229.0 & 2.89 $\pm$ 0.36 & 399 \\
 2532 - 2532b  & 860.0 & 64.64 $\pm$ 0.38 & 400 \\
 2504 - 2504b  & 186.0 & 4.17 $\pm$ 0.38 & 397 \\
 2381 - 2381b  & 302.0 & 1.24 $\pm$ 0.35 & 386 \\
 2065 - 2065b  & 682.0 & 58.90 $\pm$ 1.02 & 384 \\
\label{table:catalog1}
\end{longtable}
\tablefoot{
 For binary systems in the \textit{Multiple System} column, the first Vision Index corresponds to primary disk, i.e. the brighter one.
 }
\tablebib{(1)~\cite{van_Terwisga_2022}; (2)~\cite{Grosschedl19}; (3)~\cite{Kounkel16}.}

\FloatBarrier
\begin{table}
\caption{Stellar Masses}
\label{table:stellarmass}
\centering
\begin{tabular}{l l l l l}
\hline\hline
Multiple System & Disk Dust Mass & Disk Dust Mass & Stellar Mass & Stellar Mass
\\
  & \small{1st source} & \small{2nd source} & \small{1st source} & \small{2nd source}
  \\
(VISION Index) & (M$_{\oplus}$) & (M$_{\oplus}$) & (M$_{\odot}$) & (M$_{\odot}$)
\\
\hline\hline 
    1000 - 1006 & 11.1 $\pm$ 0.25 & 5.0 $\pm$ 0.25 & 0.372 $\pm$ 0.012 & 0.23 $\pm$ 0.005 
    \\
    1956 - 1954 & 17.6 $\pm$ 0.25 & 1.4 $\pm$ 0.25 &	0.948 $\pm$ 0.009 & - 
    \\
    2697 - 2698	& 19.3 $\pm$ 0.81 & 23.1 $\pm$ 1.0 & 0.906 $\pm$ 0.138 & -
    \\
    2784 - 2787	& 160.3 $\pm$ 0.28 & 8.8 $\pm$ 0.28 & 1.49 $\pm$ 0.345 & 0.274 $\pm$ 0.05 
    \\
    2746 - 2743	& 108.9 $\pm$ 1.12 & 31.4 $\pm$ 1.89 & - & 0.345 $\pm$ 0.055 
    \\
    2889 - 2888	& 116.4 $\pm$ 0.28 & 6.4 $\pm$ 0.28 & 0.392 $\pm$ 0.047 & - 
    \\
    2960 - 2959	& 4.6 $\pm$ 0.33 & 6.3 $\pm$ 0.33 & 0.352 $\pm$ 0.052 & 1.116 $\pm$ 0.196 
    \\
    1130 - 1107	& 29.3 $\pm$ 0.37 & 3.7 $\pm$ 0.43 & 0.238 $\pm$ 0.009 & 0.593 $\pm$ 0.005 
    \\
    2142 - 2141	& 2.0 $\pm$ 0.40 & 61.1 $\pm$ 0.37 & 0.296 $\pm$ 0.011 & 0.347 $\pm$ 0.008 
    \\
    2303 - 2306	& 1.4 $\pm$ 0.35 & 1.5 $\pm$ 0.36 & - & 0.322 $\pm$ 0.093 
    \\
    2230 - 2229	& 41.6 $\pm$ 0.39 & 3.3 $\pm$ 0.40 & 0.376 $\pm$ 0.005 & 0.293 $\pm$ 0.013 
    \\
    2418 - 2420	& 15.6 $\pm$ 0.37 & 12.4 $\pm$ 0.36 & - & 0.786 $\pm$ 0.01 
    \\
    2501 - 2502	& 42.4 $\pm$ 0.38 & 4.4 $\pm$ 0.38 & 0.473 $\pm$	0.021 & -
    \\
    2811 - 2812	& 2.9 $\pm$ 0.30 & 1.7 $\pm$ 0.30 & 0.183 $\pm$	0.05 & - 
    \\
    3101 - 3100	& 16.9 $\pm$ 0.37 & 3.6 $\pm$ 0.37 & - & 1.425 $\pm$ 0.015 
    \\
    2205 - 2209	& 4.4 $\pm$ 0.34 & 2.1 $\pm$ 0.37 & 1.378 $\pm$ 0.032 & - 
    \\
    2563 - 2565	& 102.7 $\pm$ 0.35 & 6.8 $\pm$ 0.37 & 1.198 $\pm$ 0.019 & - 
    \\
    2672 - 2680	& 12.8 $\pm$ 0.31 & 3.1 $\pm$ 0.26 & 0.556 $\pm$ 0.004 & - 
    \\
    2696 - 2701	& 45.7 $\pm$ 0.30 & 1.4 $\pm$ 0.32 & - & 1.048 $\pm$ 0.04 
    \\
    2861 - 2862 & 1.4 $\pm$ 0.30 & 20.5 $\pm$ 0.31 & - & 1.757 $\pm$ 0.042 
    \\
    2010 - 2009	& 5.0 $\pm$ 0.25 & 1.89 $\pm$ 0.33 & - & 0.444 $\pm$ 0.122 
    \\
    2314 - 2313	& 1.4 $\pm$ 0.25 & 1.12 $\pm$ 0.25 & - & 0.184 $\pm$ 0.01 
    \\
    2545 - 2544	& 23.1 $\pm$ 1.0 & 1.75 $\pm$ 0.28 & - & 0.37 $\pm$ 0.018 
    \\
    3011 - 3009	& 8.8 $\pm$ 0.28 & 14.82 $\pm$ 0.78 & - & 1.313 $\pm$ 0.388 
    \\
    2301 - 2304	& 31.4 $\pm$ 1.89 & 3.88 $\pm$ 0.29 & 1.128 $\pm$ 0.027 & - 
    \\
\hline
\end{tabular}
\tablebib{
\textit{Stellar information:}~\cite{DaRio}; \textit{Disk information:}~\cite{van_Terwisga_2022}.
}
\end{table}

\end{appendix}

\end{document}